\documentclass[a4paper,fleqn]{cas-dc}

\usepackage[numbers]{natbib}

\usepackage{amsmath,amssymb,amsfonts}
\usepackage{mathtools}
\usepackage{graphicx}
\usepackage{float}
\usepackage{textcomp}
\usepackage{hyperref}
\usepackage{textpos}
\usepackage{wrapfig}
\usepackage{booktabs}
\usepackage{longtable}
\usepackage{diagbox}
\usepackage{autobreak}
\usepackage{multirow}
\usepackage{fancyhdr}
\usepackage{siunitx}
\usepackage{subfigure}
\usepackage{placeins}
\usepackage{tabularx}
\usepackage[noend,ruled,boxed,commentsnumbered,linesnumbered]{algorithm2e}
\usepackage{epstopdf}
\usepackage{makecell}
\usepackage[table]{xcolor}
\usepackage{colortbl}
\usepackage{amsthm}

\def\tsc#1{\csdef{#1}{\textsc{\lowercase{#1}}\xspace}}
\tsc{WGM}
\tsc{QE}

\begin{document}
\let\WriteBookmarks\relax
\renewcommand{\topfraction}{0.9}
\renewcommand{\bottomfraction}{0.8}
\renewcommand{\textfraction}{0.07}
\renewcommand{\floatpagefraction}{0.8}
\renewcommand{\dbltopfraction}{0.9}
\renewcommand{\dblfloatpagefraction}{0.8}
\setcounter{topnumber}{4}
\setcounter{bottomnumber}{2}
\setcounter{totalnumber}{6}

\shorttitle{Nightjar: Dynamic Adaptive Speculative Decoding for LLM Serving}

\shortauthors{R. Li et al.}

\title [mode = title]{Nightjar: Dynamic Adaptive Speculative Decoding for Large Language Models Serving}



\author[1]{Rui Li}[auid=1,bioid=1]
\ead{lirui.r21@nudt.edu.cn}
\credit{Conceptualization, Methodology, Data curation, Writing - original draft, Writing - review \& editing}

\author[1]{Zhaoning Zhang}[auid=2,bioid=2,orcid=0009-0009-9320-2798]
\cormark[1]
\ead{zhangzhaoning@nudt.edu.cn}
\credit{Supervision}

\author[1]{Libo Zhang}[auid=3,bioid=3]
\ead{zhanglibo@nudt.edu.cn}
\credit{Visualization, Investigation}

\author[1]{Huaimin Wang}[auid=4,bioid=4]
\ead{hmwang@nudt.edu.cn}
\credit{Project administration}

\author[1]{Xiang Fu}[auid=5,bioid=5]
\ead{fuxiang13@nudt.edu.cn}
\credit{Funding acquisition}

\author[1]{Zhiquan Lai}[auid=6,bioid=6]
\ead{zqlai@nudt.edu.cn}
\credit{Writing - review \& editing}

\affiliation[1]{organization={State Key Laboratory of Complex \& Critical Software Environment, National Key Laboratory of Parallel and Distributed Computing, College of Computer Science and Technology, National University of Defense Technology},
            city={Changsha},
            state={Hunan},
            country={China}}

\cortext[cor1]{Corresponding author}


\begin{abstract}
Speculative decoding (SD) accelerates LLM inference by verifying draft tokens in parallel. However, this method presents a critical trade-off: it improves throughput in low-load, memory-bound systems but degrades performance in high-load, compute-bound environments due to verification overhead.
Existing speculative decoding methods use fixed lengths and cannot adapt to workload changes or decide when to stop speculation. The cost of restarting speculative inference also remains unquantified. Under high load, the benefit of speculation diminishes, while retaining the draft model reduces KV cache capacity, limiting batch size and degrading throughput. 
To overcome this, we propose Nightjar, a resource-aware adaptive speculative framework. It first adjusts to the request load by dynamically selecting the optimal speculative length for different batch sizes. Crucially, Nightjar proactively disables speculative decoding when the MAB planner determines that speculation is no longer beneficial, and during the disabled phase, offloads the draft model to the CPU only under GPU memory pressure. This reclaims memory for the KV cache, thereby facilitating larger batch sizes and maximizing overall system throughput.
Experiments show that Nightjar achieves up to 14.76\% higher throughput than standard speculative decoding and up to 20.18\% lower latency in the main benchmark suite under dynamic request arrival rates for real-time LLM serving scenarios.
\end{abstract}




\begin{keywords}
speculative decoding \sep large language models serving \sep resource efficient inference \sep dynamic real-time request loads
\end{keywords}

\maketitle

\section{Introduction}
Large language models (LLMs) are increasingly deployed in real-world applications, where they serve concurrent requests with varying throughput requirements. However, LLM serving suffers from high latency and low throughput due to the strictly sequential dependency of autoregressive decoding, which limits parallelism and leads to poor GPU utilization \cite{chitty2023survey, xia-etal-2024-unlocking}. 

\begin{figure}
\centering
\includegraphics[width=0.9\columnwidth]{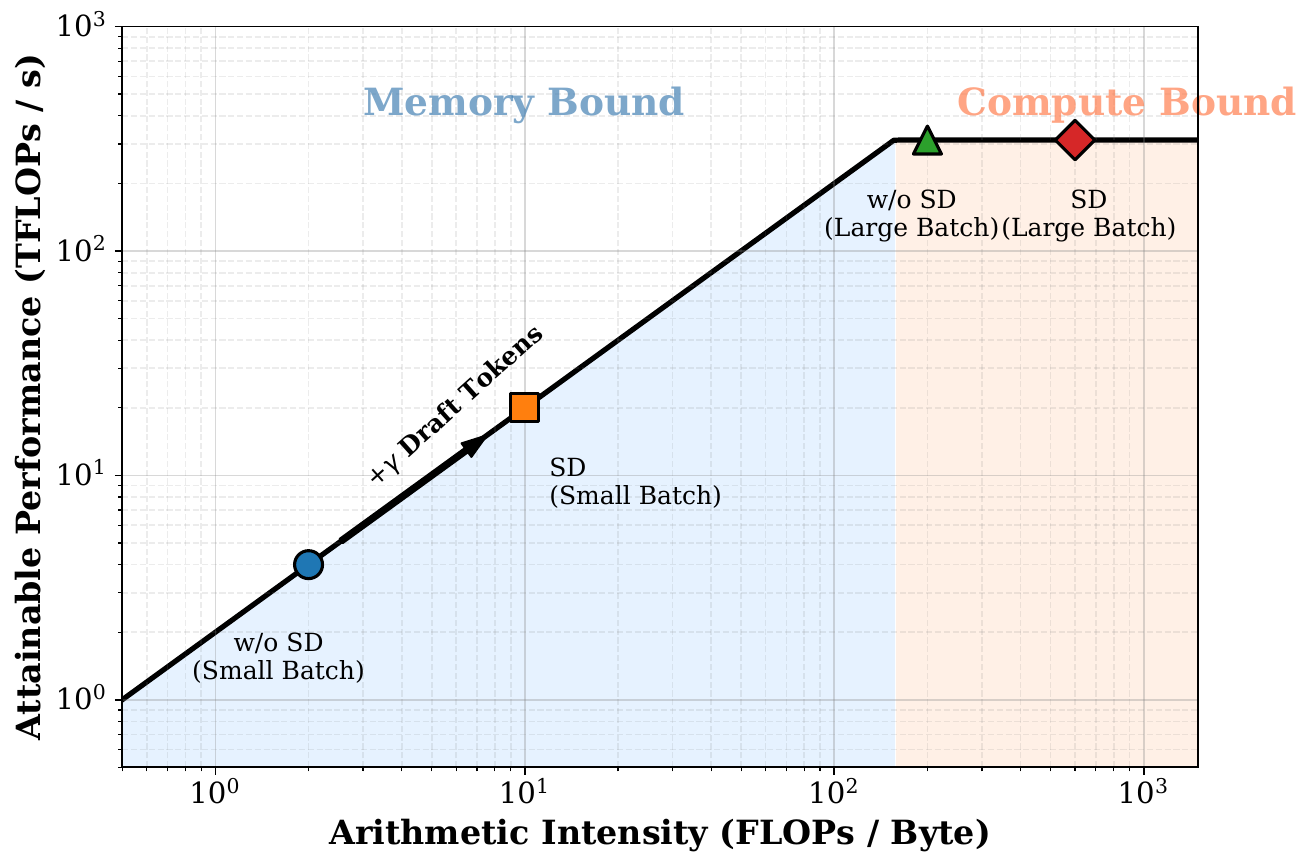}
\caption{Conceptual roofline model illustrating the performance shift in LLM inference between Vanilla Decoding (w/o SD) and Standard Speculative Decoding (SD); positions are illustrative.}
\label{fig:roofline}
\end{figure}

Speculative decoding (SD) \cite{leviathan2023fast, zhang2025dovetail} accelerates inference by allowing a draft model to propose multiple candidate tokens, while the target model verifies them in parallel, which is theoretically lossless in terms of accuracy. This reduces the strictly sequential dependency in autoregressive decoding and increases concurrency. From a roofline perspective (Figure \ref{fig:roofline}), under small batch sizes, speculative decoding raises arithmetic intensity by verifying multiple draft tokens per weight, enhancing utilization in the memory-bound regime. As batch size increases, both SD and autoregressive decoding move toward the compute-bound regime. In this region, performance is capped by the hardware’s compute ceiling, and the additional verification overhead of SD not only fails to improve throughput but can even reduce it.


However, current speculative decoding serving systems (e.g., in vLLM \cite{kwon2023efficient}) often use a fixed speculative length, which is not optimal for dynamic request loads. 
The optimal speculative length depends on batch size, token acceptance rates, and hardware characteristics. An excessively long speculative length increases the time spent on draft generation and imposes significant verification overhead, whereas a length that is too short fails to fully capitalize on the potential speedup of speculative decoding. The system should also simply disable speculation when the overhead outweighs the benefits.


Furthermore, model-based draft models consume both compute resources and GPU memory, unlike model-free approaches such as N-gram that use almost no additional memory. However, model-free methods lack generalizability and are often restricted to specific domains like code generation \cite{liu2025speculative}. Under high request load, the memory occupied by these draft models directly competes with the KV cache. Reclaiming this space for the KV cache would facilitate larger batch sizes and significantly maximize overall system throughput.

Most optimizations for speculative decoding fail to account for dynamic workloads in real-world serving. Early works targeted single requests \cite{wang_opt-tree_2025,zhang_adaeagle_2024,zhang_draft_2024,brown_dynamic_2024,huang_specdec_2024} or static batches \cite{su2023synergy, houbanditspec}, making them unsuitable for fluctuating request loads. Newer dynamic approaches \cite{su2023synergy,liu2024optimizing,huang_specserve_2025,li_adaserve_2025} also have limitations. For example, SpecServe \cite{huang_specserve_2025} and TETRIS \cite{wu2025tetris} rely on draft confidence, which scales poorly with large batch sizes. DSD \cite{liu2024optimizing} optimizes speculative length using historical acceptance rates, but it suffers from a speculation reactivation difficulty: once speculation is disabled, no new speculative observations are collected, making re-enabling harder, and it also wastes GPU memory by keeping the idle draft model resident.
Similarly, BanditSpec \cite{houbanditspec} introduces a Multi-Armed Bandit approach for hyperparameter selection, but its static design fails to incorporate real-time batch sizes as context and overlooks the heavy KV cache reconstruction penalty

Even if speculative length is chosen adaptively, resource contention for draft model weights and KV cache remains unresolved. While existing works such as Oneiros \cite{li2025oneiros} and eLLM \cite{xu2025ellm} attempt to optimize GPU memory through memory remapping or shared pools, they predominantly focus on general-purpose memory management within or across models. Consequently, they overlook the intrinsic coupling between speculative policy decisions and memory allocation, particularly the dependency between speculative length selection and draft model residency.

Therefore, we need a system that adapts to request load changes and balances costs and benefits in real time. It must be capable of dynamically selecting an appropriate speculative strategy for different batch sizes, including disabling it when necessary. To address this core challenge, we propose \textbf{Nightjar}, a contextual bandit approach for speculative length selection in dynamic, real-time LLM services. A key feature of our approach is its ability to learn an effective speculative length for different batch sizes under changing serving conditions, without requiring prior knowledge of request difficulty or model characteristics. At a system level, Nightjar seeks to maximize serving efficiency, equivalently minimizing the effective latency per committed token under dynamic load. We also consider the cost of switching the speculative length from 0 to a positive number. 
When we disable speculative decoding, we offload the draft model to free up memory for the KV cache, which allows for larger batch sizes. Later, if the request load drops, we reload the draft model in the background to speed up inference again.
Implemented on vLLM \cite{kwon2023efficient} and evaluated on three real-world datasets, Nightjar speeds up LLM inference and outperforms existing baselines, proving its effectiveness in real-world scenarios.

\section{Background}
\label{sec:background}

\subsection{LLM Serving and PagedAttention}
\label{subsec:pageattention}

Text generation in Large Language Models is an autoregressive and serial process. This means the system generates only one token at a time. The next token strictly depends on all previously generated tokens.

To avoid recomputing past data, modern LLM serving systems save the hidden states of past tokens in the GPU memory. This is called the KV cache \cite{pope2023efficiently}. To improve hardware utilization, systems batch multiple requests together. However, static batching is inefficient: since requests have varying lengths, the GPU must wait for the longest one to finish before launching the next batch.

To solve this compute waste, Orca \cite{yu2022orca} introduced continuous batching (or iteration-level scheduling).  Instead of waiting for the whole batch to finish, it dynamically evicts finished requests and inserts new ones at every decoding step. 

While continuous batching improves compute utilization, it complicates memory management. Because output lengths are unknown and requests dynamically enter and leave the batch, allocating contiguous KV cache memory becomes difficult, leading to severe memory fragmentation and limiting batch size.

To address this, vLLM \cite{kwon2023efficient} introduces PagedAttention, which partitions the KV cache into fixed-size blocks and maps logical tokens to non-contiguous physical memory. This design eliminates external fragmentation and enables efficient memory sharing and swapping under continuous batching.

\subsection{Speculative Decoding}
\label{subsec:speculative_decoding}

Even with PagedAttention solving the memory allocation issue, step-by-step token generation still faces a major performance bottleneck. Since the system only generates one token at a time, the GPU's massive computing units (ALUs) spend most of their time waiting for data to load from memory. The system is heavily memory-bound.

To break this slow serial execution and improve hardware utilization, Speculative Decoding \cite{leviathan2023fast} was proposed. It changes the traditional step-by-step generation into a  draft-then-verify pipeline. Two models work together in this process:

\begin{itemize}
    \item \textbf{Drafting:} First, the system uses a small, fast draft model to quickly guess the next $\gamma$ tokens in a row. Here, $\gamma$ is the speculative length.
    \item \textbf{Verifying:} Next, the system feeds these $\gamma$ draft tokens into the large target model all at once. The target model uses its parallel computing power to verify all $\gamma$ tokens in a single forward pass.
\end{itemize}

If the draft model guesses correctly, the system outputs multiple tokens in one step, greatly speeding up the process. If a token is guessed wrong, the target model discards it at the error point and automatically corrects it with the right token.

Through this design, speculative decoding turns a memory-bound serial process into a compute-bound parallel process. It achieves significant speedups without losing any output accuracy.

\section{Motivation}
While speculative decoding can significantly improve performance in theory, its effectiveness in real-world LLM serving is highly dependent on workload characteristics and system conditions. 
\subsection{Dynamic Speculative Decoding Length Selection}
\begin{figure}
    \centering
    \subfigure[]{
    \includegraphics[width=0.45\linewidth]{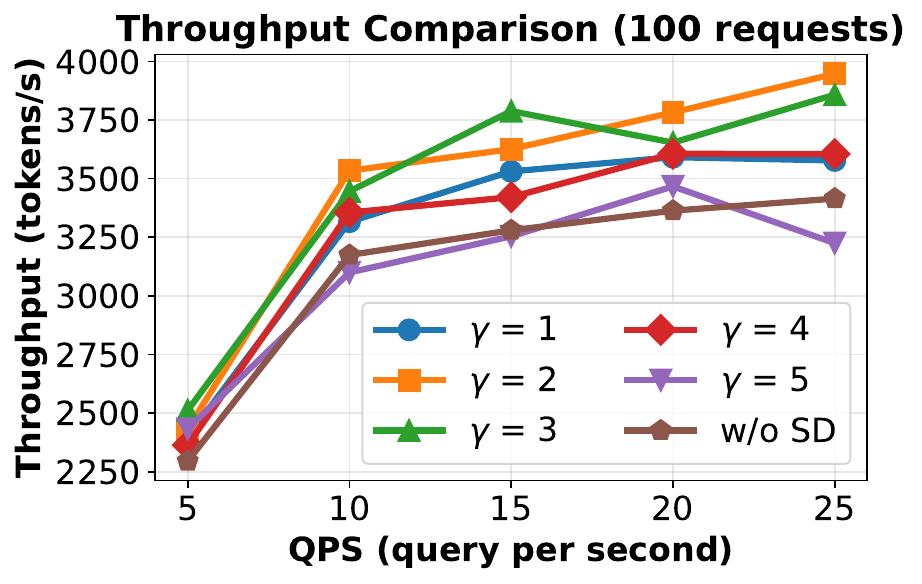}
    \label{fig:experiments_20}
    }
    \subfigure[]{
    \includegraphics[width=0.45\linewidth]{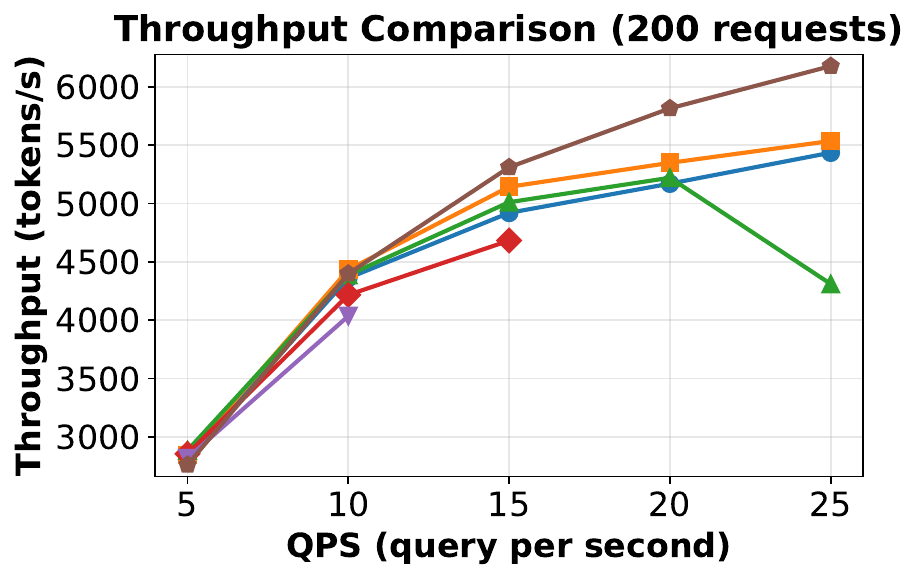}
    \label{fig:experiments_300}
    }
    \caption{Throughput under varying request rates. (a) 100 requests; (b) 200 requests. $\gamma$ denotes the speculative length (tokens generated by the draft model per step). Note: configurations with $\gamma \geq 4$ are omitted at high QPS due to GPU out-of-memory (OOM) errors.}
    \label{fig:experiments}
    \end{figure}
    In speculative decoding, the draft model generates $\gamma$ tokens, which are subsequently verified by the target model in parallel. Consequently, the total latency per decoding step is the sum of the draft model's generation time and the target model's verification time. The actual number of tokens produced per step equals the number of accepted tokens ($n$) plus one bonus token. While increasing the speculative length $\gamma$ can potentially raise the number of accepted tokens, it also incurs higher computational cost. If the acceptance rate (the ratio of accepted tokens to $\gamma$) is low, the added overhead may outweigh the gains, ultimately leading to a decrease in overall throughput.

    We empirically validate this trade-off\footnote{Experiments were conducted on the ShareGPT dataset using vLLM with the DeepSeek-R1-Distill-Qwen-7B target model and the DeepSeek-R1-DRAFT-Qwen2.5-0.5B draft model \cite{alamios_DeepSeek_R1_DRAFT_2024} on an RTX 4090 GPU.}. In this figure, the reported behavior corresponds to model-based speculative decoding under this fixed target/draft-model pairing; it is not a model-free N-gram setting. As shown in Figure~\ref{fig:experiments_20}, SD provides significant speedups at lower request loads. For instance, a speculative length ($\gamma$) of 3 improves throughput by 15.5\% at 15 QPS while a speculative length of 2 yields better improvement than 3 at QPS=25. However, as the load increases, Vanilla Decoding (w/o SD) surpasses SD's performance. At high loads (Figure \ref{fig:experiments_300}), the overhead becomes detrimental, leading to a performance degradation of up to 30.25\% compared to Vanilla Decoding (w/o SD). This confirms that SD excels in the memory-bound regime, while its verification overhead becomes detrimental when the system is compute-bound.
    
    The core challenge lies in the dynamic and unpredictable nature of real-world LLM serving request loads. Consequently, a fixed speculative length is inherently suboptimal, as a configuration tuned for low-load efficiency can degrade service under high request load. The optimal speculative length is not a static value but a dynamic variable dependent on real-time factors like system load, hardware characteristics, and token acceptance rates \cite{sadhukhan2024magicdec}.

\begin{figure}
    \centering
    \includegraphics[width=0.65\columnwidth]{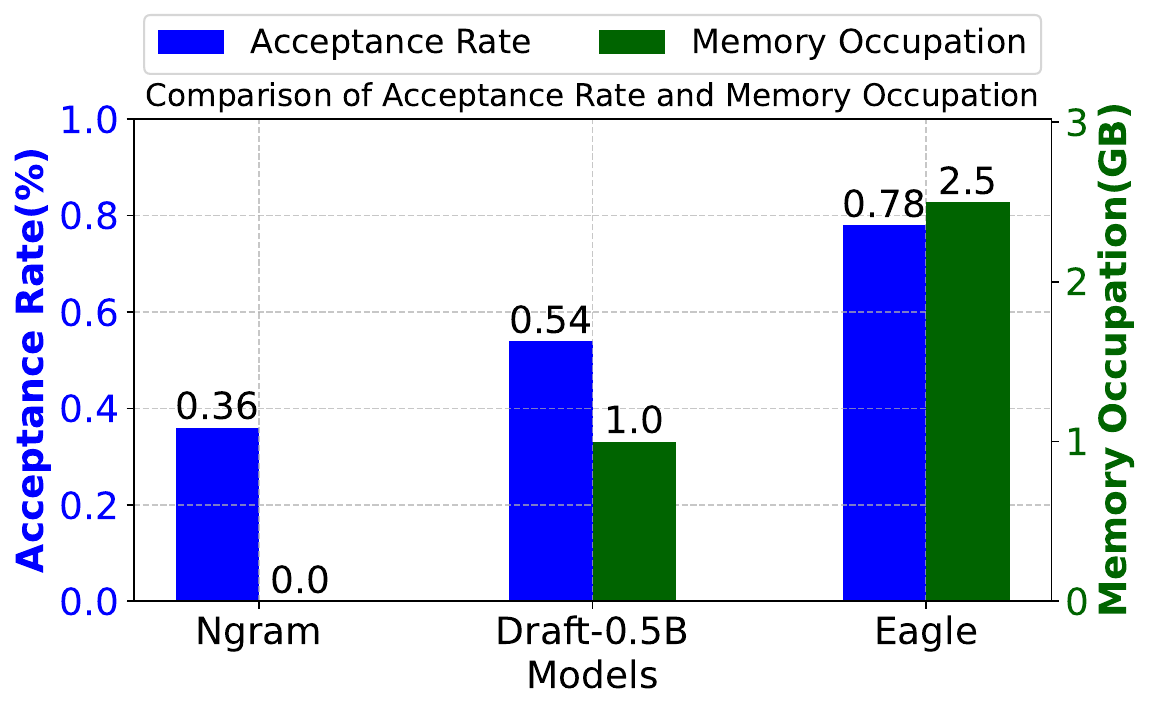}
    \caption{Memory occupation for different draft methods. The reported memory footprints are measured on an RTX 4090 GPU under the 7B serving configuration, using DeepSeek-R1-Distill-Qwen-7B as the target model. For the model-based methods, the compared drafts include DeepSeek-R1-DRAFT-Qwen2.5-0.5B and Eagle; N-gram is a model-free baseline and therefore introduces almost no additional neural-model memory footprint.}
    \label{fig:experiments_mem}
    \end{figure}

\subsection{Memory underutilization}

    \begin{table}[htbp]
        \centering
        \caption{KV cache memory occupation for different models for a single request (batch size 1).}
        \label{tab:kv-cache-comparison}
        \begin{tabularx}{\columnwidth}{lXX}
        \hline
        \textbf{Context Length} & \textbf{Vicuna-13B} & \textbf{DeepSeek-R1-Distill-Qwen-7B} \\ \hline
        1,024 (1K)               & 800 MiB                   & 56 MiB                                   \\ \hline
        4,096 (4K)               & 3,200 MiB        & 224 MiB                                    \\ \hline
        8,192 (8K)               & 6,400 MiB      & 448 MiB                                   \\ \hline
        32,768 (32K)             & 25,600 MiB     & 1,792 MiB                   \\ \hline
        \end{tabularx}
        \end{table}

        Beyond adaptive length selection, memory capacity introduces an additional system-level constraint. Table \ref{tab:kv-cache-comparison} shows that the KV cache can consume a large amount of GPU memory, and its size grows rapidly with context length.  Because memory is limited, a larger KV cache means there is less space available for the draft model's weights. This resource contention exposes a problem with static memory partitioning: as the KV cache grows, it leaves less space for draft models, which lowers system throughput.

As shown in Figure \ref{fig:experiments_mem}, draft models (N-gram, DeepSeek-R1-DRAFT-Qwen2.5-0.5B \cite{alamios_DeepSeek_R1_DRAFT_2024}, Eagle \cite{li2024eagle}) occupy varying GPU memory, impacting KV cache capacity.   While their weights are smaller than the main model, they constrain memory under high loads. This memory overhead is particularly critical for large-scale models. Consider a standard deployment of DeepSeek-V3 671B \cite{guo2025deepseek} across 10$\times$H100 (80GB) GPUs. The 14B MTP draft model, distributed via tensor parallelism, consumes 1.4GB of GPU memory per GPU, which is extremely valuable, equivalent to the storage of $\sim$49k KV cache tokens. 
Current systems (e.g., vLLM \cite{kwon2023efficient}) retain draft models even when SD is disabled, wasting memory.
    
This inefficiency stems from the rigid separation between draft model weights and the KV cache in GPU memory, preventing flexible reallocation and increasing KV cache pressure.



\section{System Architecture} \label{sec:system_arch}

\begin{figure}
    \centering
    \includegraphics[width=1.0\columnwidth]{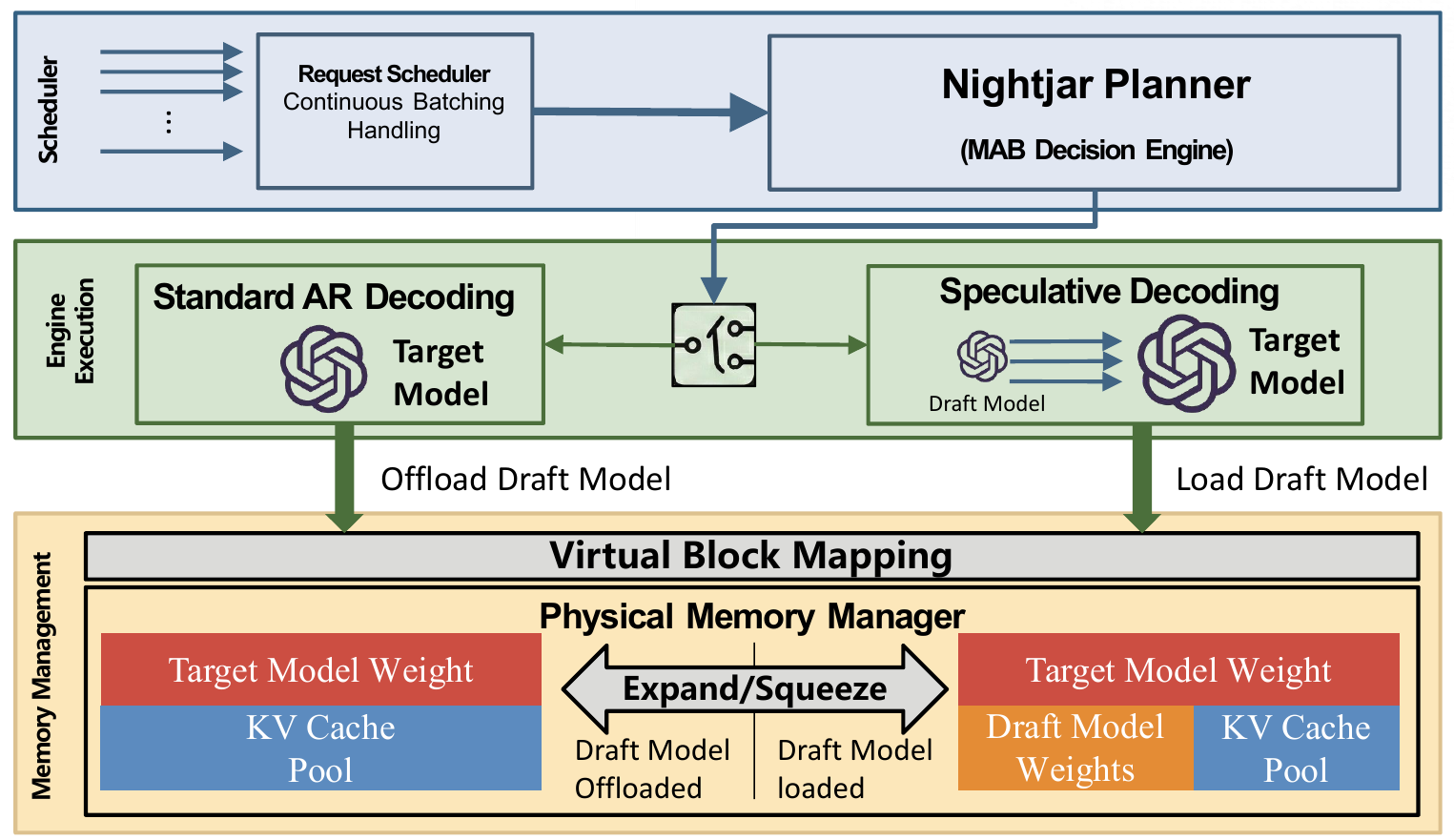}
    \caption{The Nightjar architecture operates across three layers: (1) the \textbf{Scheduler} manages incoming requests; (2) the \textbf{Planner} makes step-wise decisions to toggle speculation; (3) the \textbf{Memory Manager} dynamically reallocates resources between model weights and the KV cache pool based on system load.}
    \label{fig:overview}
\end{figure}

Figure \ref{fig:overview} shows the architecture of Nightjar. It consists of three main components that work together to make serving decisions: the Scheduler, the Planner, and the Memory Manager.

\textbf{1) Scheduler and Planner Layer.} 
At the top level, the \textit{Request Scheduler}  runs at every decode step and handles continuous batching for incoming requests. It feeds real-time batch statistics (current batch size $B$) to the \textbf{Nightjar Planner}, which employs a contextual multi-armed bandit engine to determine the optimal speculative length $\gamma$ for the current decoding step. This decision is reused within the current bin rather than being resampled at every step. 

\textbf{2) Engine Execution Layer.} 
Driven by the Planner's decision, the execution engine dynamically switches between two modes:
\begin{itemize}
    \item \textit{Speculative Decoding (Right):} When the Planner selects $\gamma > 0$, the system engages the draft model to propose tokens, which are verified in parallel by the target model.
    \item \textit{Vanilla Decoding (w/o SD) (Left):} If the Planner selects $\gamma = 0$, speculative decoding is disabled and the system reverts to standard autoregressive decoding.
\end{itemize}

\textbf{3) Memory Management Layer.} 
Underlying the execution is an elastic Memory Manager. As shown in the bottom panel of Figure \ref{fig:overview}, this module manages the contention between draft model weights and the KV cache. It implements a "Squeeze/Expand" mechanism: when the speculative decoding is disabled and  memory is insufficient for the KV cache, it offloads the draft model to host memory to expand the KV cache pool, maximizing batch size. Conversely, during low load, it reallocates memory to reload the draft model before re-enabling speculative decoding, ensuring accelerated inference. Reload restores the system's ability to speculate, but the Planner may still keep $\gamma=0$ until speculation again becomes favorable.

This design helps Nightjar handle workload changes by adjusting both algorithms and system resources.

\section{Multi-Armed Bandit for Speculative Length Selection}
In this section, we introduce the multi-armed bandit (MAB) for speculative length selection.
\subsection{Problem Statement}
In continuous batching \cite{kwon2023efficient} for LLM serving, the batch size can vary at each decoding step. We consider batch sizes $B \in \mathcal{B} = \{1, 2, \dots, B_{\max}\}$ and speculative lengths $\gamma \in \{0, 1, \dots, \Gamma_{\max}\}$. 
While performance is often evaluated using \textbf{goodput} $g_{B,\gamma}$ (tokens/s) \cite{liu2024optimizing}. Here, $g_{B,\gamma}$ denotes the average number of committed tokens per decoding step divided by the corresponding step latency, where the committed tokens include all successfully verified draft tokens plus one additional target-model bonus token. Directly maximizing goodput ignores the heavy latency overhead caused by KV cache reconstruction during strategy switches. 

To explicitly account for switching overhead, we formulate the problem in terms of the \emph{effective} latency per token.
At time $t$, the algorithm plays arm $\gamma_t$ and observes the latency per token $\ell_t(\gamma_t)$ (the inverse of the goodput at that step). 
The \emph{empirical mean} latency for batch size $B$ and arm $\gamma$ is $\tilde{\ell}_{B,\gamma} = \frac{1}{N_{B,\gamma}} \sum_{s: \gamma_s=\gamma, B_s=B} \ell_s(\gamma)$, where $N_{B,\gamma}$ is the number of steps that selected $\gamma$ under batch size $B$ (the sum runs over \emph{observed} latencies at those steps).

Since the token acceptance rate is unknown a priori, we model this as a regret minimization task over $T$ steps. Let $\gamma_t$ denote the speculative length chosen at step $t$.
We define the \textbf{Loss} observed at step $t$, denoted as $L_t(\gamma_t)$, to be the realized latency per token plus the switching overhead:
\begin{equation}
L_t(\gamma_t) = \ell_t(\gamma_t) + \mathbb{I}(\gamma_{t-1}=0 \land \gamma_t>0) \cdot \frac{C_{\text{switch}}}{\gamma_t}
\end{equation}
where $\mathbb{I}(\cdot)$ is the indicator function and $C_{\text{switch}}$ is the latency overhead of KV cache reconstruction amortized over the post-switch tokens.

This overhead corresponds to the KV cache reconstruction cost incurred when re-enabling speculative decoding. 
When the system transitions from $\gamma=0$ to $\gamma>0$, the draft model, which remained inactive during autoregressive decoding, must perform an additional prefill pass to reconstruct the missing KV cache states. 
This introduces a one-time system-level latency. 
We amortize this cost by dividing it by $\gamma_t$, which serves as a lightweight proxy for the post-switch token yield. 
Since the selected arm is locked within each bin (Algorithm~\ref{alg:nightjar}), the one-time overhead is naturally spread over subsequent decoding steps.

We compare against the best fixed arm in hindsight under the same batch size sequence. Formally, for each batch size $B$, let $\mu_{B,\gamma}$ denote the mean latency per token for arm $\gamma$ when the batch size is $B$. We define
\begin{equation}
\gamma^*(B) = \arg\min_{\gamma \in \{0,\dots,\Gamma_{\max}\}} \mu_{B,\gamma}.
\end{equation}
The cumulative regret over horizon $T$ is
\begin{equation}
    \begin{aligned}
    R(T) 
    = \sum_{t=1}^T \Big(
    & \ell_t(\gamma_t) \\
    & + \mathbb{I}(\gamma_{t-1}=0 \land \gamma_t>0) 
      \cdot \frac{C_{\text{switch}}}{\gamma_t} \\
    & - \ell_t(\gamma^*(B_t))
    \Big)
    \end{aligned}
    \end{equation}

Our goal is to find an arm that minimizes $R(T)$. We summarize the key notations and their descriptions in Table \ref{tab:notations1}.

\begin{table}[htbp]
    \centering
    \caption{Summary of key notations used in Nightjar for speculative length selection.}
    \label{tab:notations1}
    \begin{tabularx}{\columnwidth}{|l|X|} 
    \hline
    \textbf{Symbol} & \textbf{Description} \\ \hline
    
    \rowcolor[gray]{0.9} \multicolumn{2}{|l|}{\textit{System Parameters \& Cost Model}} \\ \hline
    $B, B_{\max}$ & Current and maximum batch size, where $B \in \{1, \dots, B_{\max}\}$. \\ \hline
    $\gamma, \Gamma_{\max}$ & Current and maximum speculative length (draft tokens). \\ \hline
    $\delta_{\max}$ & Effective skip length, defined as $\max_{1 \le i \le B} \delta_i$ (max lag in a batch). \\ \hline
    $C_{\text{switch}}$ & Constant representing the KV cache reconstruction cost in the loss function. \\ \hline

    \rowcolor[gray]{0.9} \multicolumn{2}{|l|}{\textit{MAB Hierarchical Structure}} \\ \hline
    $t$ & Global decoding time step index. \\ \hline
    $j_B$ & Block index for batch size $B$, governing long-term scheduling scales. \\ \hline
    $b_B$ & Bin index within block $j_B$, controlling exploration probability ($p \propto 1/b_B$). \\ \hline
    $\tau_B$ & Round counter within bin $b_B$. \\ \hline
    $H_B$ & Duration of the current block, growing exponentially as $2^{j_B-1}$. \\ \hline

    \rowcolor[gray]{0.9} \multicolumn{2}{|l|}{\textit{Performance Metrics \& Regret Analysis}} \\ \hline
    $g_{B,\gamma}$ & Goodput (tokens/s) under batch size $B$ and speculative length $\gamma$. \\ \hline
    $\ell_t(\gamma_t)$ & Observed latency per token at time $t$ when arm $\gamma_t$ is played (only this is observed). \\ \hline
    $\tilde{\ell}_{B,\gamma}$ & Empirical mean latency, $\frac{1}{N_{B,\gamma}}\sum_{s:\gamma_s=\gamma,B_s=B} \ell_s(\gamma_s)$, used for arm selection. \\ \hline
    $\mu_{B,\gamma}$ & Mean latency per token for batch size $B$ and arm $\gamma$. \\ \hline
    $\gamma^*(B)$ & Optimal speculative length for batch size $B$ (minimizes $\mu_{B,\gamma}$). \\ \hline
    $L_t(\gamma_t)$ & Realized loss at step $t$ with speculative length $\gamma_t$, including latency and switching penalty. \\ \hline
    $R(T)$ & Cumulative regret (in terms of latency) over horizon $T$. \\ \hline
    \end{tabularx}
\end{table}

\subsection{Algorithm Design}
Contextual MAB can take dynamic request batch size or input sequences as context, such as linear Thompson \cite{agrawal2013thompson} or LinUCB \cite{li2010contextual}, which have been successfully applied in online prediction settings like configuration selection \cite{stolcke2023adaptive}. However, their complexity makes them questionable for latency-sensitive dynamic serving environments. 
We adopt the relatively simple framework ADA-BINGREEDY proposed in \cite{luo2018efficient}. Its bin-level locking is particularly suitable here because it suppresses action oscillation and amortizes real switching costs, whereas simple linear contextual models are less robust to the strongly nonlinear reward surface of speculative serving. A key difference between \cite{luo2018efficient} and our approach is that we develop a request batch mechanism to handle varying request patterns, and we design a \textbf{loss function} that explicitly considers the speculative length switching cost.

\begin{figure}
\centering
\includegraphics[width=0.99\linewidth]{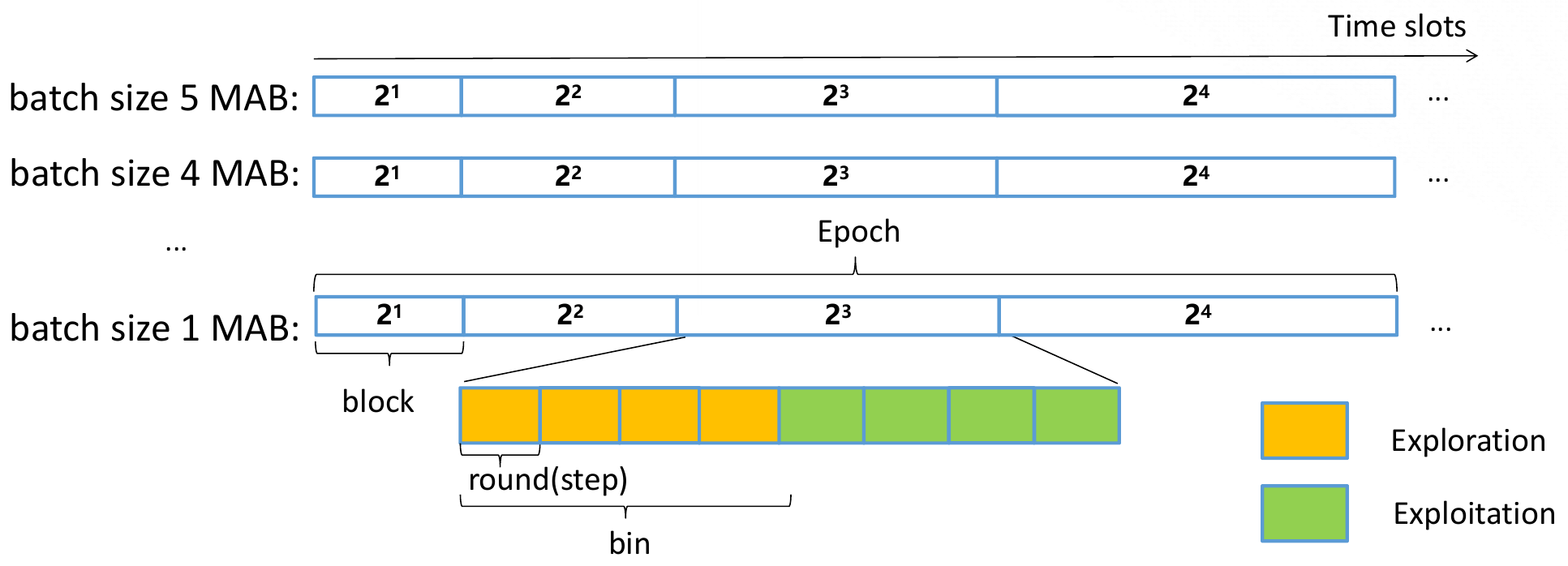}
\caption{Illustration of the Nightjar algorithm's three-level hierarchy. 
Each batch size maintains its own independent timeline. 
This timeline is organized into \textbf{Blocks ($j_B$)} of exponentially growing duration, which govern the long-term scheduling behavior.
These blocks are further divided into fixed-size \textbf{Bins ($b_B$)}, the basic units for controlling the exploration-exploitation trade-off. Each \textbf{round ($\tau_B$)} is a decoding step.}
\label{fig:algo_structure}
\end{figure}

As shown in Figure \ref{fig:algo_structure}, the Nightjar algorithm adaptively selects the speculative length $\gamma$ for each batch under continuous batching by organizing time into blocks and bins to balance exploration and exploitation. 

\begin{figure}
\centering
\includegraphics[width=0.92\linewidth]{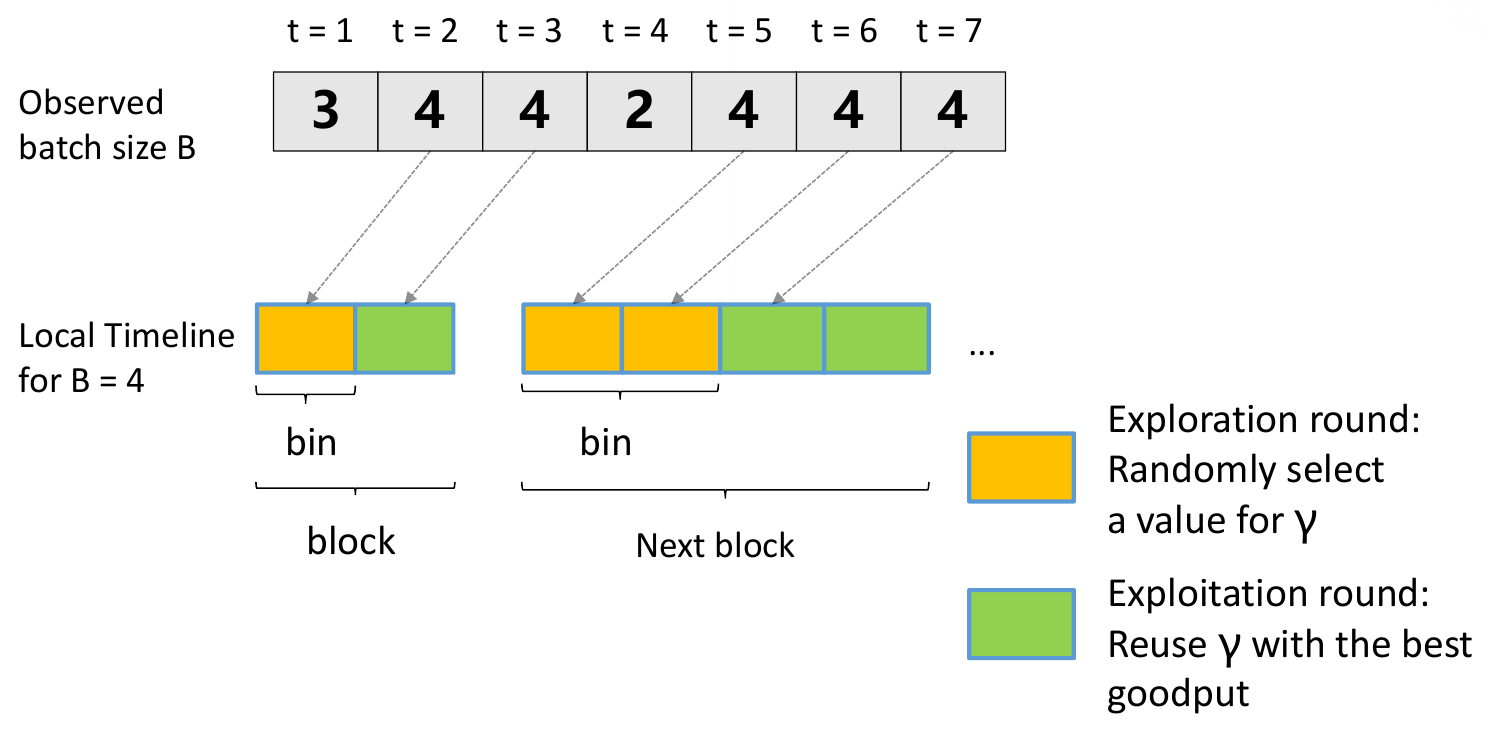}
\caption{A concrete example of the hierarchy under continuous batching. The top row shows the global batch-size trace, while the bottom row shows the local timeline for $B=4$. Orange rounds denote exploration and green rounds denote exploitation.}
\label{fig:algo_trace}
\end{figure}

Figure \ref{fig:algo_trace} gives a concrete short-trace example. If the observed batch sizes are $(3,4,4,2,4,4,4)$, then only the steps with batch size $4$ are counted on the local timeline of $B=4$, yielding five local rounds. These local rounds are grouped into bins and then blocks; within one bin, the selected $\gamma$ is kept fixed and is updated only when a new bin starts. An epoch is a higher-level logical phase on this same per-batch-size local timeline.

To explicitly account for the switching overhead, we reformulate the maximization of goodput into the minimization of latency per token.
Formally, at global time $t$ with batch size $B$, the selected speculative length $\gamma_t$ for an exploitation step is determined by minimizing the effective latency:

\begin{equation}
    \gamma_t = \mathop{\arg\min}\limits_{\gamma \in \{0,1,\dots,\Gamma_{\max}\}}  \left\{ 
        \tilde{\ell}_{B,\gamma} + \frac{\mathbb{I}(\gamma_{t-1}=0 \land \gamma>0) \cdot C_{\text{switch}}}{\gamma} \right\}
\label{eq:exploitation_min}
\end{equation}

The second term captures the KV cache reconstruction overhead incurred when re-enabling speculative decoding.

Algorithm~\ref{alg:nightjar} details the execution flow. To handle the asynchronous nature of continuous batching, Nightjar maintains independent state variables for each batch size $B$, organizing time into a hierarchical structure of \textit{blocks} (indexed by $j_B$) and \textit{bins} (indexed by $b_B$). 

\noindent\textbf{Initialization and Bin Selection.} For every time step $t$, the algorithm identifies the current batch size $B$. Critical decision-making occurs at the onset of a new bin (when the round counter $\tau_B=1$). At this step, the algorithm determines the bin type: it is designated as an \textit{exploration} bin with probability $1/b_B$ and as an \textit{exploitation} bin otherwise. This decaying exploration probability encourages more exploration in early stages and gradually shifts toward exploitation as $b_B$ increases.

\noindent\textbf{Arm Selection and Execution.}  If it is an exploration bin, $\gamma$ is sampled uniformly from the available search space to ensure coverage (Lines~\ref{line:explore_start}--\ref{line:explore_end}). Conversely, in an exploitation bin, the algorithm selects the optimal $\gamma$ by solving Eq.~\eqref{eq:exploitation_min} (Lines~\ref{line:exploit_start}--\ref{line:exploit_end}). For all subsequent steps within the bin, the system executes this fixed $\gamma$, observes the realized goodput $g_t$, and updates the internal counters. This locking mechanism is essential for bounding the switching cost.

A key feature of Nightjar is its exponential time scaling. A bin concludes when the round counter $\tau_B$ exceeds $\sqrt{H_B}$, and a block concludes when the bin counter $b_B$ exceeds $\sqrt{H_B}$ (Lines~\ref{line:hier_start}--\ref{line:hier_end}). Upon the completion of a block, the block size is updated via $H_B \gets 2^{j_B-1}$. This exponential growth ensures that as the history length $j_B$ increases, the duration of stable exploitation phases grows significantly, allowing the system to capitalize on refined goodput estimates.

\noindent\textbf{Prefill Cost Modeling.}
The effectiveness of Eq.~\eqref{eq:exploitation_min} relies on an accurate estimation of $C_{\text{switch}}$, 
which represents the KV cache reconstruction latency. This cost is not static; 
it varies based on the hardware state and the synchronization gap between the draft and target models. 

\begin{table}[htbp]
  \centering
  \caption{Measured switching cost ($C_{\text{switch}}$) across varying batch sizes and input lengths for DeepSeek-R1-Distill-Qwen-7B \cite{deepseek2024deepseekr1}, evaluated on an NVIDIA RTX 4090.}
  \label{tab:ttft_diff}
  \begin{tabular}{ccc}
    \hline
    \textbf{Input Length} & \textbf{Batch Size} & \textbf{$C_{\text{switch}}$ (ms)} \\
    \hline
    128 & 32 & 17.87 \\
    128 & 64 & 28.53 \\
    \hline
    256 & 32 & 20.65 \\
    256 & 64 & 22.33 \\
    \hline
    512 & 32 & 24.30 \\
    512 & 64 & 102.03 \\
    \hline
\end{tabular}
  \end{table}

We quantify this overhead using the \textit{effective skip length}, defined as $\delta_{\max} = \max \{\delta_1, \dots, \delta_B\}$, 
where $\delta_i$ is skip length for request $i$, defined as the number of
tokens for which the draft model has not computed KV cache
during the disabled speculation phase and must re-prefill
when speculation is re-enabled. The algorithm queries an offline-populated lookup table $C_{\text{switch}}(\delta_{\max}, B)$ (Table \ref{tab:ttft_diff}) to obtain the precise cost. 
To construct this table, we performed offline profiling across a grid of input lengths and geometrically increasing batch sizes (e.g., 2, 4, 8, \dots). Specifically, we recorded the prefill time for the standalone target model ($T_{\text{base}}$) and the prefill time with speculative decoding enabled ($T_{\text{SD}}$). The switching cost is derived from the difference $C_{\text{switch}} = T_{\text{SD}} - T_{\text{base}}$. At runtime, unseen $(\delta_{\max}, B)$ pairs are estimated by interpolation over nearby profiled points, and the resulting table is deployment-specific rather than hardware-agnostic. As shown in Table \ref{tab:ttft_diff}, on an RTX 4090 setup, this cost can range from 17.87 ms to over 102.03 ms, making dynamic retrieval essential.

\begin{algorithm}[t]
    \caption{Nightjar algorithm.}
    \label{alg:nightjar}
    \SetKwInOut{Input}{Input}
    \SetKwInOut{Initialize}{Initialize}
    \Input{Max speculative length $\Gamma_{\max}$, Max batch size $B_{\max}$}
    \Initialize{Global time $t \gets 1$}
    \For{$B = 1$ \textbf{to} $B_{\max}$}{
        \tcp{Initialize per-batch hierarchy states} 
        $j_B \gets 1$, $H_B \gets 1$, $b_B \gets 1$, $\tau_B \gets 1$ 
    }
    \For{each time step $t=1,2,\ldots$}{
        Receive current batch size $B=B_t$ \\
        \If{$\tau_B = 1$} { \label{line:bin_check}
            \tcp{Bin Start: Select Strategy \& Arm}
            Let $p = 1/b_B$ \\
            Sample $u \sim \text{Uniform}(0,1)$ \\
            \If{$u < p$}{ \label{line:explore_start}
                \tcp{Exploration: Random Arm}
                $\gamma_{\text{curr}} \sim \text{Uniform}(\{0, \dots, \Gamma_{\max}\})$ \label{line:explore_end}
            }
            \Else{ \label{line:exploit_start}
                \tcp{Exploitation: Best Arm (Lock for Bin)}
                $\gamma_{\text{curr}} \leftarrow \text{Eq.}~\eqref{eq:exploitation_min}$ \label{line:exploit_end}
            }
        }
        
        \tcp{Execute the locked arm for this step}
        Play $\gamma_t \leftarrow \gamma_{\text{curr}}$ \\
        Observe and update statistic $\tilde{\ell}_{B,\gamma_t}$ \label{line:play} \\
        $\tau_B \gets \tau_B + 1$ \\
        
        \If{$\tau_B > \sqrt{H_B}$} { \label{line:hier_start}
            \tcp{Bin Completed}
            $b_B \gets b_B + 1$, $\tau_B \gets 1$ \\
            \If{$b_B > \sqrt{H_B}$} { 
                \tcp{Block Completed}
                $j_B \gets j_B + 1$, $H_B \gets 2^{j_B-1}$, $b_B \gets 1$ \label{line:hier_end}
            }
        }
        $t \gets t + 1$ 
    }
\end{algorithm}

\textbf{Overhead.} The arm selection execution time is approximately 1e-5 seconds, while the average generation time per token in the LLM is 0.034 seconds, a difference of nearly 3400 times. This minimal overhead demonstrates the exceptional efficiency of our method, as the cost of dynamic arm selection is negligible compared to the intrinsic latency of token generation.

\subsection{Theoretical Guarantees for System Stability}While dynamic speculative decoding introduces the risk of system thrashing (i.e., frequent KV cache reconstruction), Nightjar's hierarchical bin-locking mechanism mathematically bounds this overhead. By restricting strategy switches to bin boundaries, the total number of switches is tightly bounded. As detailed in Appendix~\ref{sec:appendix_proof}, we formulate the cumulative regret $R(T)$ in terms of the system's token latency. Theoretical analysis proves that Nightjar achieves a sublinear cumulative regret of $R(T) = \tilde{\mathcal{O}}(\sqrt{T})$. This guarantees that the system rapidly converges to the optimal hardware-aware speculative length without suffering from linear performance degradation caused by strategy switching.

\section{Elastic Memory Management for Model-Based Draft Models}

\begin{figure}
  \includegraphics[width=\columnwidth]{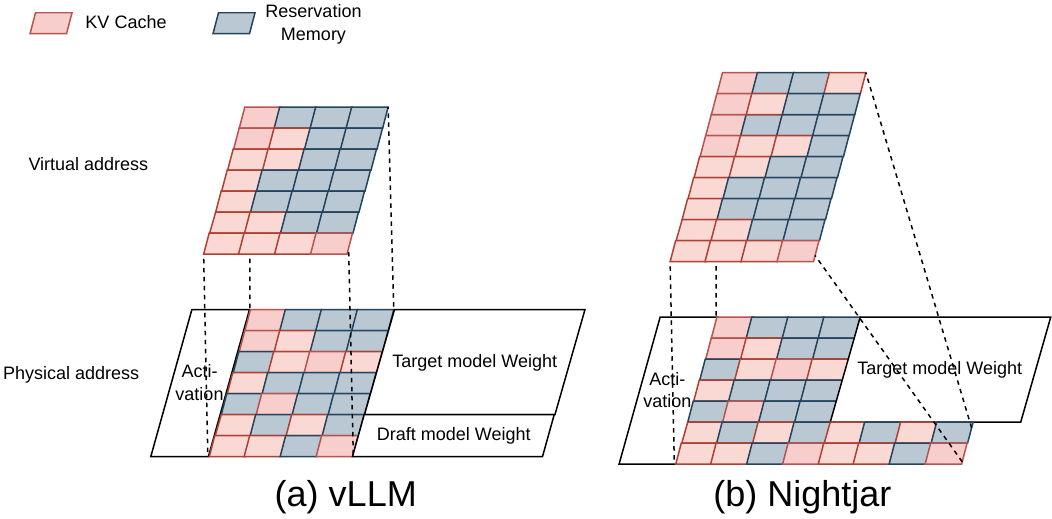}
  \caption{(a) vLLM virtualizes the KV cache but separates model weights and KV cache memory. 
  (b) Nightjar logically unifies them in a shared memory pool, enabling dynamic reallocation.}
  \label{fig:visualize_nightjar_offload}
\end{figure}

When request load is high, KV cache memory becomes constrained, leading to frequent swaps or request drops. In this setting, the draft model occupies valuable GPU memory, further reducing the space available for the KV cache and worsening system throughput. To address this, we propose a dynamic offloading mechanism for model-based draft models.  As illustrated in Figure \ref{fig:visualize_nightjar_offload}, we offload the draft model weights to the host memory, analogous to LoRA adapter management, thereby reclaiming GPU memory to expand the capacity of the original KV cache. Both KV cache and draft model sizes are measured in bytes for consistency. Let $M_{\text{orig}}$ denote the baseline KV cache memory size (bytes); the expanded memory size is $M_{\text{scale}} = M_{\text{orig}} + S_{\text{draft}}$, where $S_{\text{draft}}$ is the draft model weight memory size (bytes). Equivalently, in block count: $N_{\text{scale}} = N_{\text{orig}} + N_{\text{draft}}$ with $N_{\text{draft}} = \lceil S_{\text{draft}} / B_{\text{block}} \rceil$ and $B_{\text{block}}$ the bytes per KV cache block. Symmetrically, when system load decreases, the KV cache is compressed back to its baseline size, and the draft model weights are reloaded to GPU memory to resume speculative decoding. Both KV cache expansion and contraction are performed asynchronously with respect to ongoing decoding and do not block the generation pipeline.

\begin{table}[htbp]
  \centering
  \caption{Summary of key notations used in Nightjar for elastic memory management.}
  \label{tab:notations2}
  \begin{tabularx}{\columnwidth}{|l|X|} 
  \hline
  \textbf{Symbol} & \textbf{Description} \\ \hline
   $M_{\text{orig}}, M_{\text{scale}}$ & Baseline and expanded KV cache memory size (bytes); $M_{\text{scale}} = M_{\text{orig}} + S_{\text{draft}}$. \\ \hline
   $S_{\text{draft}}$ & Draft model weight memory size (bytes). \\ \hline
   $B_{\text{block}}$ & Bytes per KV cache block. \\ \hline
   $N_{\text{orig}}, N_{\text{scale}}, N_{\text{draft}}$ & KV cache capacity in blocks: baseline, expanded, and draft-equivalent ($N_{\text{draft}} = \lceil S_{\text{draft}}/B_{\text{block}}\rceil$). \\ \hline
   $N_{\text{free}}$ & Number of currently available physical KV cache blocks. \\ \hline
   $\tau_{\text{low}}$ & Critical scarcity threshold (in blocks) triggering memory expansion. \\ \hline
   $T_{\text{persist}}$ & Temporal window size for validating low-memory states. \\ \hline
   $|\mathcal{Q}_{\text{wait}}|$ & Cardinality of the request waiting queue. \\ \hline
  \end{tabularx}
\end{table}

  \subsection{Trigger Conditions for Dynamic Adjustment}

The system employs an elasticity policy based on real-time KV cache block availability and request queue status. A \textbf{KV cache expansion} event is triggered when the following conditions are all met: 
(1) Speculative decoding is disabled and the draft model weights have been successfully offloaded from GPU memory; 
(2) The number of available memory blocks, denoted as $N_{\text{free}}$, falls below a critical threshold $\tau_{\text{low}}$ ($N_{\text{free}} < \tau_{\text{low}}$); 
(3) The low-memory state persists for a continuous window of $T_{\text{persist}}$ scheduling steps. 

Conversely, a \textbf{KV cache contraction} is initiated to reload the draft model when: 
(1) The waiting queue is empty ($|\mathcal{Q}_{\text{wait}}| = 0$); 
(2) Sufficient free blocks exist to accommodate both the draft model and a safety buffer ($N_{\text{free}} > N_{\text{draft}} + \tau_{\text{low}}$). 
In practice, contraction is drain-aware: it does not require the expanded region to be completely drained. It can proceed once reclamation no longer requires moving a large amount of live KV state, and can otherwise be deferred until more blocks become drainable.
This hysteresis design ensures that speculative decoding remains disabled during high-load phases (where $N_{\text{free}} < \tau_{\text{low}}$), thereby preventing resource thrashing. In implementation, the reload condition is also required to persist for multiple consecutive steps, so transient low-load fluctuations do not trigger immediate reload.
We summarize the key notations and their descriptions in Table \ref{tab:notations2}.

\subsection{Non-blocking Asynchronous Migration}

Once an expansion decision is made, the system first offloads the draft model weights to host memory. During this phase, draft model weights are immediately evicted from GPU memory, while a backup copy remains in host memory. The reclaimed GPU memory is then reassigned to the KV cache pool. Conversely, when contraction is triggered, the system reloads the draft model to GPU memory after the KV cache is shrunk to its original capacity.

Both offloading and reloading are executed using asynchronous CUDA streams with non-blocking Direct Memory Access (DMA). This design enables data transfer to overlap with ongoing computation, ensuring that memory management operations incur negligible impact on the serving latency of the primary model.

\subsection{Mechanism of KV cache expansion}

Upon triggering expansion, the system dynamically reallocates GPU memory. The Model Executor allocates an additional physical memory region of size $S_{\text{draft}}$, which is logically attached to the KV cache pool. Subsequently, the Block Manager updates its internal state: 
(1) The set of allocatable physical block indices is expanded to include the new range $[K_{\text{boundary}}, K_{\text{total}})$; 
(2) Reference counters for these newly instantiated blocks are initialized to zero; 
(3) The new block indices are appended to the free block queue. 
Crucially, this process does not require relocating existing KV blocks,
thereby avoiding large-scale memory copying and minimizing reconfiguration overhead.

\subsection{KV cache contraction and Logical Remapping}

\begin{figure*}
  \centering
  \includegraphics[width=\textwidth]{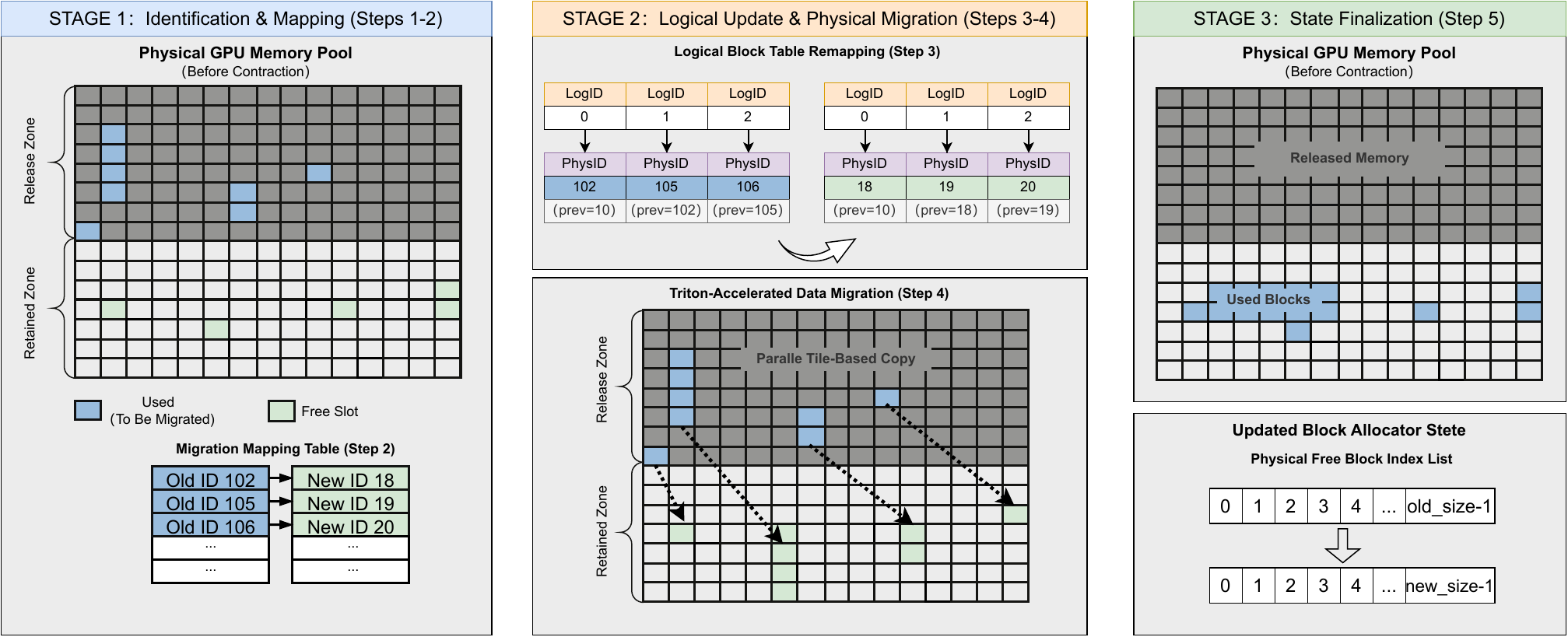}
  \caption{Overview of the KV cache block contraction and logical table remapping process. The system ensures data consistency during physical memory compaction through vectorized data migration.}
  \label{fig:kvcache_contraction}
\end{figure*}

Contraction requires reclaiming the extended memory region while ensuring that active sequences remain uninterrupted.  As depicted in Figure \ref{fig:kvcache_contraction}, the process involves five sequential steps:

\textbf{Step 1: Identification of Eviction Candidates.} 
The system scans the page tables of all active sequences to identify the set of physical blocks $\mathcal{B}_{\text{evict}}$ located in the memory region to be reclaimed (i.e., blocks with $\text{ID} \geq K_{\text{boundary}}$). These blocks must be migrated to the preserved memory region $[0, K_{\text{boundary}})$.

\textbf{Step 2: Construction of Migration Plan.} 
The system verifies that the free list contains a sufficient number of slots whose physical block IDs are strictly less than $K_{\text{boundary}}$. A one-to-one mapping $\mathcal{M}: b_{\text{old}} \rightarrow b_{\text{new}}$ is then constructed, assigning each block in $\mathcal{B}_{\text{evict}}$ to an available slot in the preserved region $[0, K_{\text{boundary}})$. This mapping serves as the basis for the subsequent metadata update and data migration phases.
Blocks selected for migration are first marked as reserved and removed from the allocator’s free list, preventing new allocations from targeting them. The compaction process then proceeds asynchronously on a separate CUDA stream.

Since decoding accesses KV blocks exclusively through logical mappings, and metadata updates are applied atomically before physical reclamation, ongoing generation is unaffected. Newly arriving requests are allocated only from the preserved region, ensuring safe concurrency between allocation, decoding, and migration.

\textbf{Step 3: Metadata Update and Remapping.} 
For each block requiring migration, the system performs atomic metadata updates: 
(a) A new physical block object is instantiated, inheriting the token sequence and size of the original; 
(b) Predecessor pointers are updated, if a predecessor block is also part of the migration set, the pointer is redirected to its new location; otherwise, the original reference is maintained; 
(c) The logical Block Table is updated to map the sequence to the new physical block IDs; 
(d) The reference counter for the new block is incremented, and it is marked as allocated. All block table remapping operations are protected by a global allocator-level synchronization barrier, ensuring that no decoding thread observes partially updated mappings.  
Finally, the old physical block is released, ensuring the logical sequence remains consistent with the physical memory state.

\textbf{Step 4: Vectorized Data Migration.} 
Following the metadata update, the physical data transfer is executed according to the mapping $\mathcal{M}$. The migration is powered by a custom high-performance kernel (implemented in Triton) utilizing a tile-based parallelization strategy. Each thread block manages the transfer of a complete KV cache block via vectorized load/store instructions.  By decoupling memory reclamation from model execution, we avoid potential deadlock scenarios caused by hold-and-wait resource patterns. When contraction is triggered, only a small number of active sequences typically reside in the extended region. As a result, the number of blocks requiring migration is limited. Combined with our Triton-optimized vectorized migration kernel, the overall overhead remains negligible.

\textbf{Step 5: Allocator State Finalization.} 
Post-migration, the block allocator trims its index set, removing IDs $\geq K_{\text{boundary}}$ and clearing associated reference counts, effectively finalizing the contraction.

\subsection{System Correctness and Consistency Guarantee}
While Nightjar optimizes resource allocation through dynamic model offloading and contextual bandit scheduling, it mathematically preserves the strong consistency of speculative decoding. The system guarantees that the final output distribution is strictly equivalent to that of the standalone target model. We ensure this absolute correctness through two key architectural designs:

\textbf{State Consistency during Dynamic Switching.} 
In Nightjar, the scheduling and execution are rigidly synchronized at a step-level granularity. The generation process alternates cleanly between standard autoregressive decoding and speculative decoding. When the MAB planner decides to disable speculative decoding ($\gamma \to 0$) for the next step, the system seamlessly transitions back to the standard autoregressive mode. In this state, the draft model does not generate any speculative tokens, and consequently, no parallel verification is invoked. Because the verification of any preceding speculative step is strictly finalized before the new step commences, there are no leftover unverified tokens. The target model simply generates the next token autoregressively, ensuring mathematical equivalence at every step boundary.

\textbf{Transparency and Atomicity of Physical Memory Migration.}
The dynamic memory management, including KV cache page table remapping and physical memory migration via Triton kernels, is transparent to the upper execution engine. Crucially, Nightjar structurally isolates these memory operations from the speculative execution phase. Physical memory movements are strictly triggered only when speculative decoding is entirely disabled. 
Specifically, \textit{KV cache expansion} occurs exclusively when $\gamma=0$ and the draft model has already been evicted from the GPU memory. Similarly, \textit{KV cache contraction} (reclaiming memory for the draft model) is executed before speculative decoding is re-enabled, while the draft model is still absent from the GPU memory. Because these physical data migrations happen solely during the pure autoregressive generation phase, the target model's attention mechanism always accesses a secure, uncorrupted, and mathematically consistent historical KV cache. This temporal isolation guarantees that memory compression and movement never interfere with the correctness of the generated outputs.

\section{Experiments}
\subsection{Experiment setup}


\begin{figure}
\centering
\includegraphics[width=0.95\linewidth]{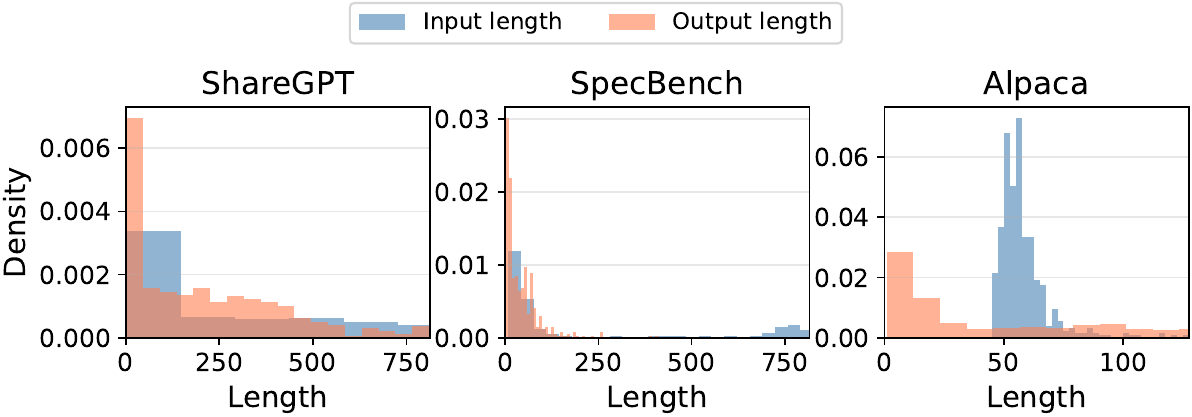}
\caption{Distribution of input and output lengths across three datasets.}
\label{fig:distributed}
\end{figure}
\noindent \textbf{Settings.} We select DeepSeek-R1-Distill-Qwen-7B \cite{deepseek2024deepseekr1} and Vicuna-13B \cite{vicuna13b} as target models and match them with  DeepSeek-R1-DRAFT-Qwen2.5-0.5B \cite{alamios_DeepSeek_R1_DRAFT_2024} and vicuna-68m \cite{double7_vicuna68m} as draft models respectively. The model pair for the 7B target model is operated on a single RTX 4090 (24GB) GPU, and the model pair for the 13B target model operates on an A100 (40GB) GPU.
 Nightjar is implemented by extending the vLLM \cite{kwon2023efficient}. In our implementation, we set $T_{\text{persist}} = 3$ to avoid reacting to short-term fluctuations.
 
\noindent \textbf{Workloads.}  To simulate
a dynamic user environment, we generate request arrivals
using a Poisson distribution. We also selected real-world production traces from the Azure LLM Inference Dataset \cite{AzurePublicDataset}, and chose a segment with dynamic request rates as shown in Figure \ref{fig:trace1}. For online chatting, we utilize datasets from ShareGPT \cite{anon8231489123_ShareGPT_V3}, Alpaca \cite{tatsu-lab_alpaca}. For other tasks,  we utilize SpecBench \cite{xia-etal-2024-unlocking}, which is composed of 480 randomly
selected instances from six widely used databases. For each dataset, we use 480 instances. The input and output length distribution is presented in Figure \ref{fig:distributed}.

\noindent \textbf{Baselines}. We evaluate the total token throughput (tokens/s) and mean end-to-end latency of Nightjar against baseline methods: 
\begin{itemize}
    \item \textbf{Standard Speculative Decoding (SD):} Implemented on top of vLLM, this baseline employs a vanilla chain-style draft generation ($\gamma=3$), where the draft model generates a sequence of candidate tokens followed by parallel verification from the target model.
    \item \textbf{Vanilla Decoding (w/o SD):} This baseline does not use speculative decoding, where the target model directly generates the response tokens without using the draft model.
    \item \textbf{Dynamic Speculative Decoding (DSD) \cite{liu2024optimizing}:} This method utilizes a linear regression model to predict execution latency and estimates the expected acceptance rate based on historical statistics. It dynamically selects the speculative length by computing goodput as the ratio of the expected number of accepted tokens (accepted length $+1$) to the predicted execution time, and choosing the length that maximizes this value.

    \item \textbf{BanditSpec \cite{houbanditspec}:} This approach treats the selection of speculative length as a Multi-Armed Bandit problem, leveraging a variant of the Upper Confidence Bound (UCB) algorithm for optimization. However, it does not explicitly incorporate batch size as a contextual feature in its decision-making process.

    \item \textbf{TETRIS \cite{wu2025tetris}:} Given a fixed speculative length (similar to standard SD), TETRIS keeps the draft generation process unchanged but redesigns the verification phase. Within a predefined computational budget, it prioritizes the verification of tokens with the highest predictive entropy across a batch to improve verification efficiency.
\end{itemize}

    
 Our result is averaged over five independent runs.
\begin{figure}
    \centering
    \includegraphics[width=0.9\linewidth]{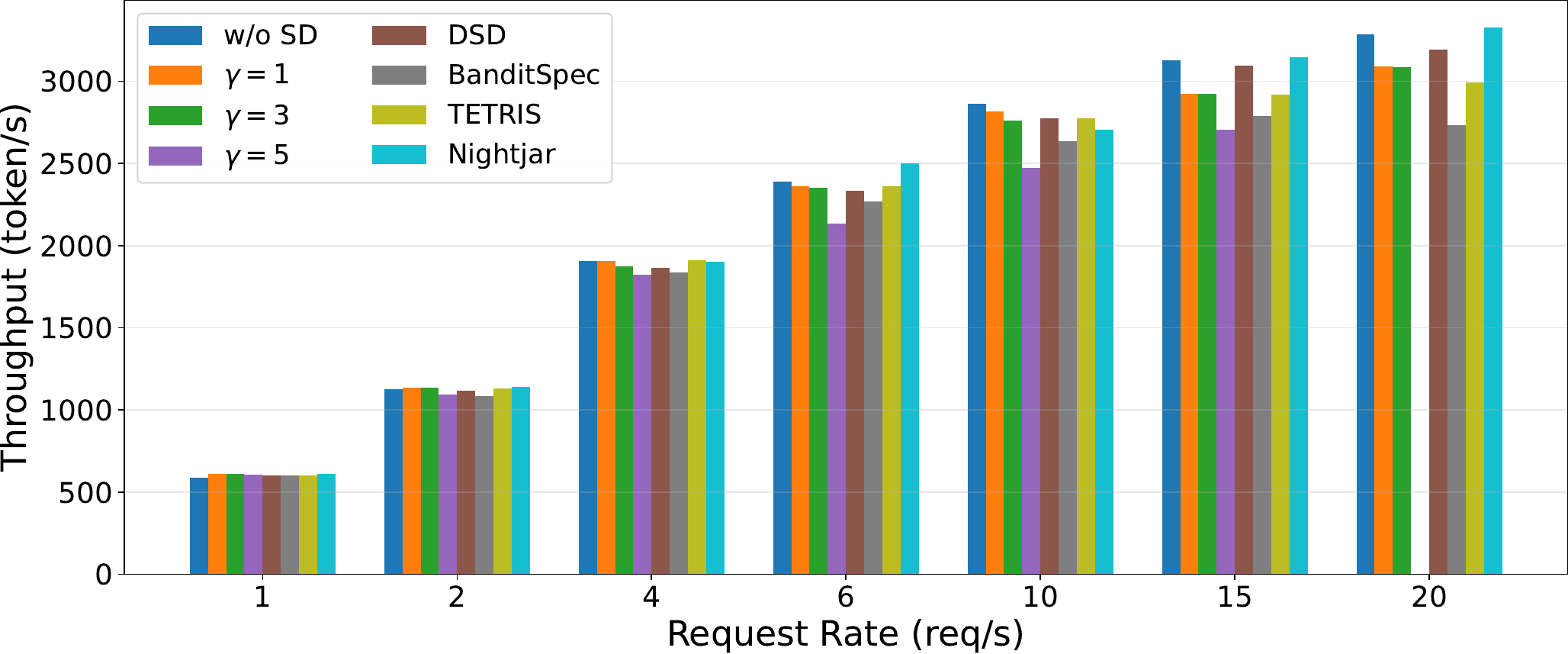} 
    \caption{Method comparison at low request rate and high request rate for 7B model.}
    \label{fig:highqps_lowqps_comparison}
    \end{figure}

\section{Experiment Results}
\subsection{Static request rate}
Nightjar maintains high throughput under different loads by dynamically adjusting its speculation strategy. 
As shown in Figure \ref{fig:highqps_lowqps_comparison}, in the detailed breakdown by request rate on ShareGPT dataset, under low request rates (Request Rate < 4) where speculative decoding  outperforms standard decoding, 
Nightjar maximizes efficiency by selecting the optimal draft length. Conversely, under high request rates where speculative decoding performs worse than the baseline due to its overhead, Nightjar disables speculation to prevent performance degradation, achieving strong performance across both low- and high-load settings.
Note that data for $\gamma=5$ at 20 QPS is excluded from Figure \ref{fig:highqps_lowqps_comparison} owing to GPU memory limitations.
\subsection{Dynamic request rate}

\begin{figure}
\centering
\includegraphics[width=0.5\linewidth]{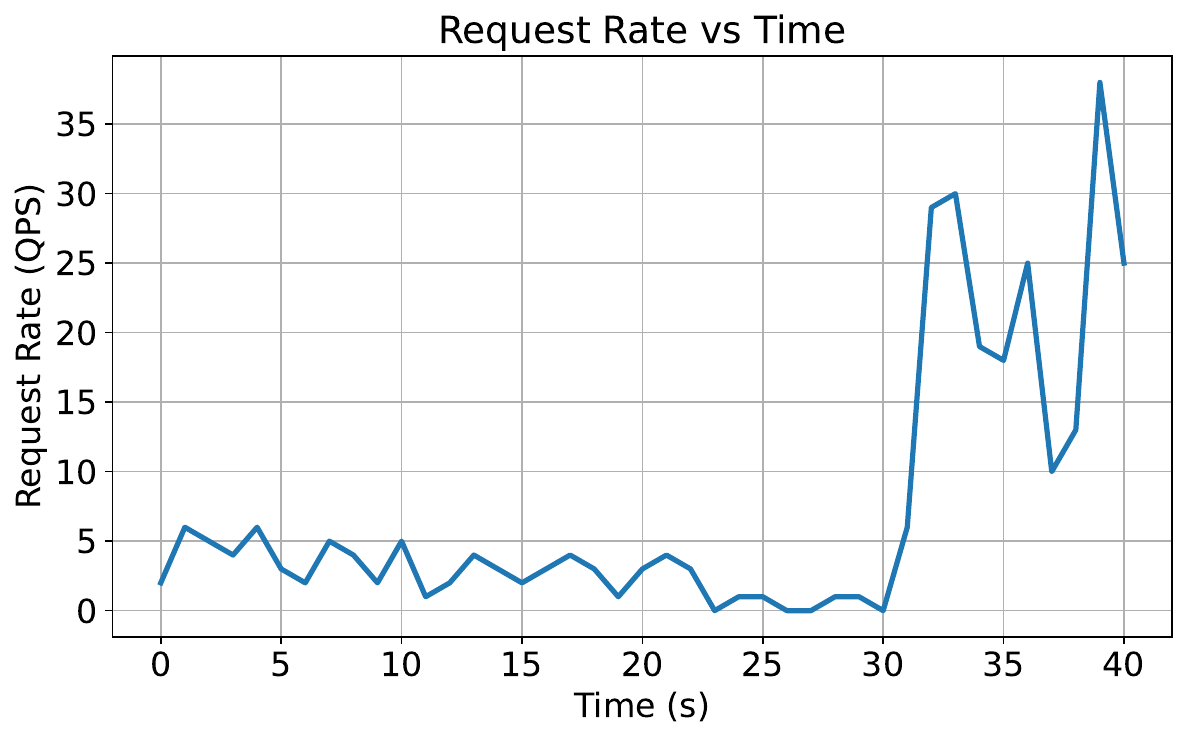}
\caption{Request rate trace.}
\label{fig:trace1}
\end{figure}

\begin{figure}
\centering
\includegraphics[width=0.9\linewidth]{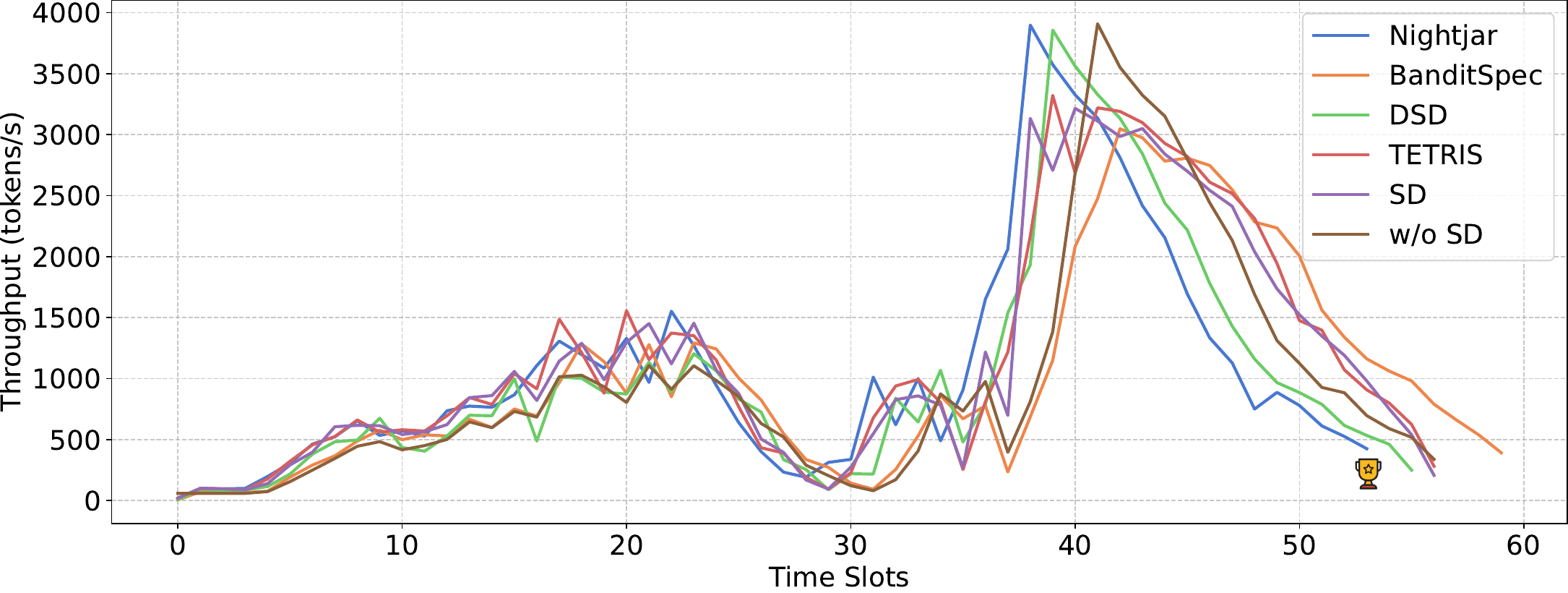}
\caption{Throughput trace for 7B model under dynamic request rate.}
\label{fig:trace}
\end{figure}
For the 7B model under dynamic request rate (Figure \ref{fig:trace1}), as shown in Figure \ref{fig:trace}, during low request load periods, Nightjar achieves the highest peak throughput by dynamically selecting the optimal speculative length. Conversely, during high-load phases, it strategically disables speculation and reverts to Vanilla Decoding (w/o SD). This prevents the severe performance degradation that plagues other speculative methods, allowing Nightjar to maintain consistently superior overall throughput across varying request load conditions. DSD's predictions are less accurate during low request load, leading to throughput degradation, but perform better under high request load, while TETRIS exhibits performance drops under high request load due to its high overhead in large batch size and BanditSpec is not suitable for dynamic batch size.

\begin{table}[htbp]
    \centering
    \caption{Throughput (tokens/s) comparison for 7B and 13B models.}
    \label{tab:throughput_comparison}
    \resizebox{\columnwidth}{!}{ 
    \begin{tabular}{c|ccc|ccc}
    \hline
    & \multicolumn{3}{c}{7B } & \multicolumn{3}{c}{13B} \\
    \cline{2-7}
    \textbf{Method} & \textbf{Alpaca} & \textbf{ShareGPT} & \textbf{SpecBench} & \textbf{Alpaca} & \textbf{ShareGPT} & \textbf{SpecBench} \\
    \hline
    w/o SD     & 2072.38 & 3653.61 & 3207.28 & 438.13 & 1290.04 & 673.48  \\
    SD         & 1920.63 & 3593.36 & 3044.80 & 628.47 & 1689.53 & 964.02 \\
    BanditSpec & 1860.74 & 3428.75 & 2631.43 & 679.82 & 1383.81 & 763.15 \\
    DSD        & 2044.40 & 3614.65 & 2789.51 & 603.32 & 1244.96 & 692.01  \\
    TETRIS     & 1875.91 & 3597.21 & 2974.14 & 702.65 & 1628.86 & 985.40 \\
    Nightjar (ours)   & \textbf{2102.73} & \textbf{3734.23} & \textbf{3326.20} & \textbf{721.23} & \textbf{1729.68} & \textbf{1061.82} \\
    \hline
    \end{tabular}
    }
    \end{table}

\begin{table}[htbp]
\centering
\caption{Mean end-to-end latency (ms) comparison for 7B and 13B models.}
\label{tab:e2elatency_comparison}
\resizebox{\columnwidth}{!}{ 
\begin{tabular}{c|ccc|ccc}
\hline
& \multicolumn{3}{c}{7B } & \multicolumn{3}{c}{13B} \\
\cline{2-7}
\textbf{Method} & \textbf{Alpaca} & \textbf{ShareGPT} & \textbf{SpecBench} & \textbf{Alpaca} & \textbf{ShareGPT} & \textbf{SpecBench} \\
\hline
w/o SD     & 8876.14  & 6938.47  & 7654.05  & 32222.49 & 9350.87& 23831.74 \\
SD         & 11110.68 & 7047.49  & 8567.04  & 24671.40 & 8152.97 & 14912.97\\
BanditSpec & 12340.83 & 7799.63  & 9288.23  & 29077.32& 9667.04 & 19294.42\\
DSD        & 9183.21  & 6499.12  & 8301.02  & \textbf{21212.44} & 10057.72& 22361.71 \\
TETRIS     & 11734.19 & 7409.62  & 8789.48  & 22133.43 & \textbf{8046.25}&14937.37\\
Nightjar (ours)    & \textbf{8868.17}  & \textbf{6534.76}  & \textbf{7618.85} & 21525.49 & 8133.92 & \textbf{14691.81} \\
\hline
\end{tabular}
}
\end{table}

As summarized in Table \ref{tab:throughput_comparison}, Nightjar delivers the best throughput among the compared methods across the six main 7B/13B benchmark settings. Averaged across the six main 7B/13B benchmark settings, our method improves throughput by 27.29\% over Vanilla Decoding (w/o SD). Relative to dynamic speculative baselines, it surpasses DSD and BanditSpec by 22.89\% and 19.76\% on average, respectively. Compared with Standard Speculative Decoding (SD), Nightjar improves throughput by 8.32\% on average across the six main 7B/13B benchmark settings and by up to 14.76\% on 13B Alpaca.

As shown in Table \ref{tab:e2elatency_comparison}, Nightjar also provides favorable end-to-end latency across the six main 7B/13B benchmark settings, reducing mean latency by 15.16\% on average relative to Vanilla Decoding (w/o SD), and by up to 20.18\% relative to Standard Speculative Decoding (SD). For the representative 13B cases where Nightjar achieves higher throughput with a small E2E-latency penalty, Appendix~\ref{sec:appendix_13b_case} shows that Nightjar improves throughput by serving more requests with larger or more efficient decoding batches, but a subset of requests experiences longer prefill/queueing delay, which slightly raises mean end-to-end latency. 


\subsection{Ablation and Analysis}

\subsubsection{Bandit Methods}

\begin{figure*}[pos=t]
    \centering
    \noindent
    \begin{minipage}[t]{0.32\textwidth}
        \centering
        \subfigure[Alpaca]{\includegraphics[width=\linewidth]{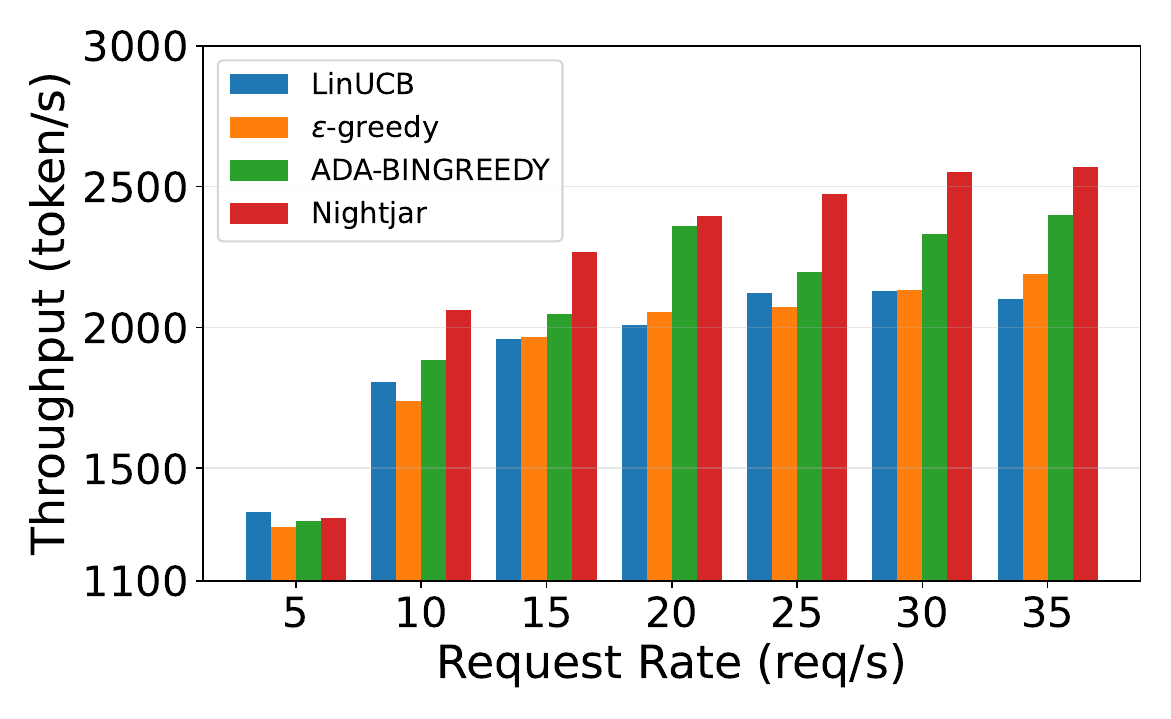}\label{fig:bandit_alpaca}}
    \end{minipage}\hfill
    \begin{minipage}[t]{0.32\textwidth}
        \centering
        \subfigure[ShareGPT]{\includegraphics[width=\linewidth]{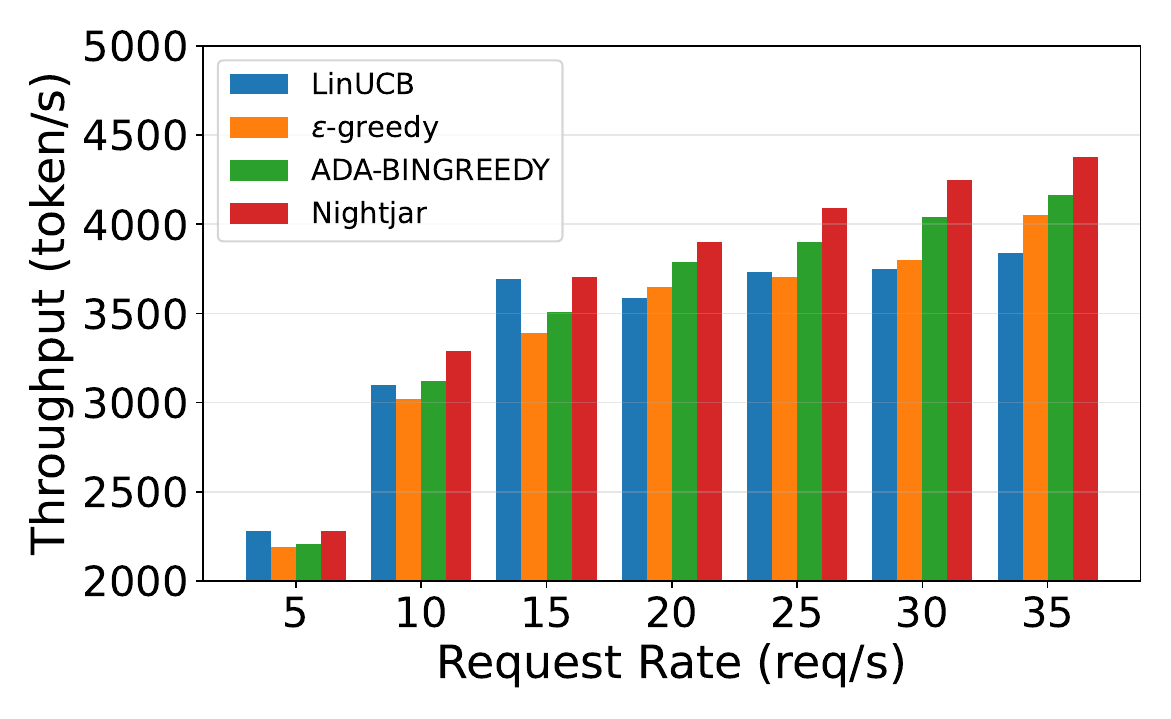}\label{fig:bandit_sharegpt}}
    \end{minipage}\hfill
    \begin{minipage}[t]{0.32\textwidth}
        \centering
        \subfigure[SpecBench]{\includegraphics[width=\linewidth]{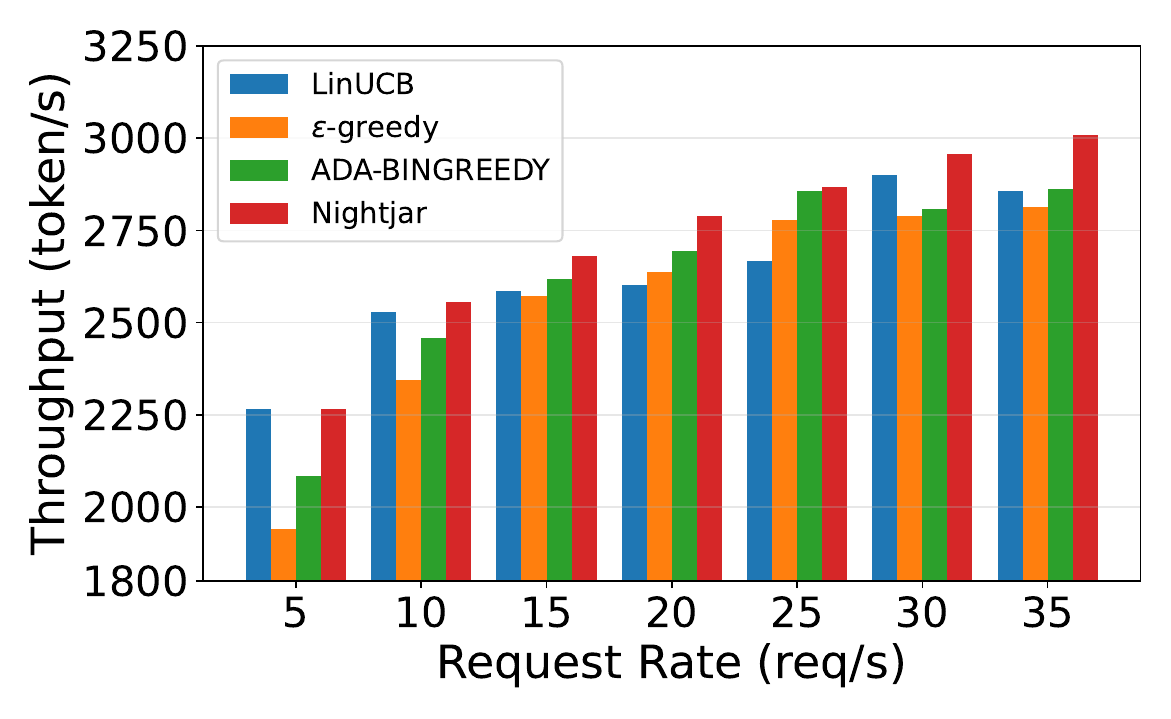}\label{fig:bandit_specbench}}
    \end{minipage}
    \caption{Context-aware bandit method comparison on Alpaca, ShareGPT and SpecBench datasets.}
    \label{fig:2}
\end{figure*}

We conduct ablation experiments on 7B model. Figure~\ref{fig:2} additionally compares epsilon-greedy, context-aware multi-armed bandit methods (with batch size as context), and the ADA-BINGREEDY baseline. Our method consistently achieves optimal performance across different request rate levels. LinUCB assumes linear relationships between rewards and contexts, which don't hold for LLM speculative serving scenarios. Throughput depends not only on hardware but also on acceptance rates and model types, making linear models unsuitable for this complex scenario. 

Compared with ADA-BINGREEDY, Nightjar consistently achieves better performance across varying workloads. This improvement highlights the importance of explicitly modeling the switching cost $C_{\text{switch}}$, which is ignored in the original ADA-BINGREEDY formulation.

\begin{table*}[htbp]
\centering
\caption{ShareGPT medium-load per-step runtime statistics. Throughput is the end-to-end system serving metric reported by the benchmark. Goodput is computed  as the average per-step committed-token throughput  rather than from the ratio of the separately reported mean committed tokens/step and mean step latency. ``Committed tokens/step'' averages the committed token count over decode steps, and ``Step latency'' averages per-step decode latency.}
\label{tab:response_acceptance_sharegpt_med}
\begin{tabular}{l|rrrrrr}
\hline
\textbf{Method} & \textbf{Throughput} & \textbf{Goodput} & \textbf{Acceptance rate} & \textbf{Mean $\gamma$} & \textbf{Committed/step} & \textbf{Step latency (ms)}\\
\hline
SD  & 1623.3 & 487.4  & 0.656 & 3.00 & 26.0 & 61.12 \\
DSD & 1641.1 & 679.9  & 0.528 & 1.64 & 25.3 & 39.21 \\
BanditSpec & 1675.2  & 746.0 & \textbf{0.690} & 2.52 & 30.9 & 45.32 \\
TETRIS & 1660.5 & 686.9 & 0.625 &  3.00  & \textbf{35.5}  & 52.13 \\
Nightjar & \textbf{1729.8} & \textbf{784.2}  & 0.667 & 2.43 & 28.0 & \textbf{37.18}  \\
\hline
\end{tabular}\par
\vspace{0.05cm}
{\small Throughput and goodput are reported in tokens/s.}
\end{table*}

To better explain how Nightjar's dynamic speculative length selection improves throughput, we next analyze per-step runtime statistics under a representative fixed-load setting (10 req/s). Unlike the dynamic-trace experiment above, this analysis uses a static offered load and is intended to expose the mechanism behind the observed throughput differences. As shown in Table~\ref{tab:response_acceptance_sharegpt_med}, Nightjar achieves the highest throughput and goodput while also yielding the lowest mean step latency. Its acceptance rate remains competitive with the strongest speculative baselines, which indicates that the gain does not come from an unusually conservative policy that simply disables useful speculation. Instead, Nightjar benefits from selecting a moderate speculative length that preserves acceptance quality while reducing verification overhead, thereby improving effective per-step efficiency.

\subsubsection{Efficacy of Dynamic Model Offloading}

\begin{figure}[pos=h]
    \centering
    \includegraphics[width=0.95\linewidth]{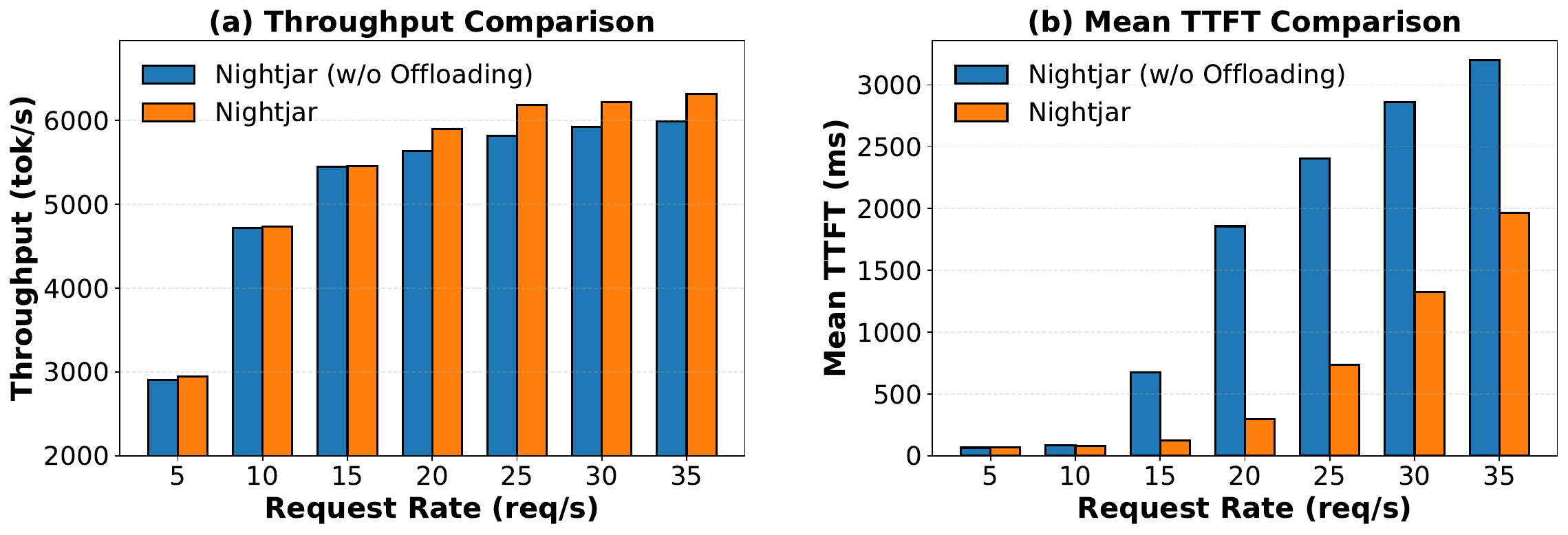}
    \caption{Performance comparison of Nightjar and Nightjar (w/o offload).}
    \label{fig:offload_comparison}
    \end{figure}


For the 7B model, the impact of our dynamic loading and offloading mechanism is assessed by comparing Nightjar with Nightjar (w/o offload). Results in Figure~\ref{fig:offload_comparison} demonstrate that offloading becomes increasingly beneficial as the system enters the compute-bound regime at high request rates, reaching a peak throughput of 6315.9 tokens/s at 35 req/s, a substantial improvement over the 5982.6 tokens/s achieved by Nightjar (w/o offload).

In addition to throughput gains, offloading also improves the average time-to-first-token (TTFT) by 47.2\% (Figure~\ref{fig:offload_comparison} ). This reduction is mainly attributed to the expanded KV cache capacity after releasing the draft model weights, which allows the scheduler to admit more requests without triggering memory pressure or swap operations. As a result, prefill execution can start earlier, reducing queuing delay and improving first-token latency under high load. These results confirm that freeing GPU memory by offloading draft models effectively expands the KV cache capacity for the target model, translating not only to higher service concurrency but also to improved responsiveness.

\begin{table*}[pos=t]
\centering
\caption{Controlled fairness ablation under the same hardware budget.}
\label{tab:fairness_ablation}
\small
\setlength{\tabcolsep}{5pt}
\begin{tabular}{l ccc}
\toprule
Method & \multicolumn{3}{c}{Tok/s / Mean E2E (ms)} \\
\cmidrule(lr){2-4}
 & 1 req/s & 5 req/s & 10 req/s \\
\midrule
\multicolumn{4}{l}{\textbf{ShareGPT}} \\
SD                   & 247.63 / 3111.65 & 1179.26 / 4501.51 & 1902.96 / 7845.07 \\
DSD                  & 247.49 / 3217.29 & 1168.51 / 4575.76 & 2158.25 / 7733.15 \\
BanditSpec           & 246.99 / 3295.07 & 1167.04 / 5294.95 & 1905.25 / 12257.89 \\
TETRIS               & 246.83 / 4327.34 & 1152.00 / 6683.42 & 2027.79 / 10609.52 \\
Nightjar (w/o offload) & 247.96 / 3088.74 & 1182.37 / 4478.93 & 2161.42 / 7492.16 \\
Nightjar             & \textbf{248.21 / 3061.28} & \textbf{1184.12 / 4459.35} & \textbf{2164.88 / 7428.54} \\
\midrule
\multicolumn{4}{l}{\textbf{Alpaca}} \\
SD                   & 47.97 / 3020.38 & 229.35 / 3915.60 & 429.08 / 5334.04 \\
DSD                  & 47.88 / 3180.50 & 229.16 / 4084.76 & 424.34 / 5521.23 \\
BanditSpec           & 47.92 / 3288.30 & 223.05 / 4900.58 & 400.09 / 8154.61 \\
TETRIS               & 47.83 / 4080.08 & 225.12 / 5642.79 & 406.93 / 8287.42 \\
Nightjar (w/o offload) & 48.08 / 2994.83 & 229.92 / 3890.71 & 431.15 / 5288.63 \\
Nightjar             & \textbf{48.22 / 2938.44} & \textbf{230.41 / 3846.92} & \textbf{433.76 / 5204.57} \\
\midrule
\multicolumn{4}{l}{\textbf{SpecBench}} \\
SD                   & 273.41 / 3284.93 & 1305.68 / 5001.60 & 2376.60 / 7839.67 \\
DSD                  & 273.22 / 3365.63 & 1289.38 / 5008.76 & 2333.93 / 9161.06 \\
BanditSpec           & 273.12 / 3393.12 & 1282.55 / 6082.61 & 1952.63 / 15864.94 \\
TETRIS               & 272.41 / 4508.26 & 1270.44 / 7310.77 & 2082.02 / 14016.29 \\
Nightjar (w/o offload) & 273.76 / 3206.41 & 1308.91 / 4971.85 & 2391.64 / 7794.58 \\
Nightjar             & \textbf{274.08 / 3188.72} & \textbf{1312.44 / 4926.14} & \textbf{2408.73 / 7718.36} \\
\bottomrule
\end{tabular}
\end{table*}

Table~\ref{tab:fairness_ablation} compares Nightjar with and without elastic offloading under the same hardware budget. Across ShareGPT, Alpaca, and SpecBench, \emph{Nightjar (w/o offload)} consistently stays ahead of the strongest static-memory baseline, showing that adaptive control alone already brings stable gains. Enabling offload further improves throughput and latency in most settings, indicating that elastic memory management provides an additional system-level benefit on top of the control policy.

To better explain the benefit of dynamic offloading, we further examine the burst-spike stress trace from the perspective of speculation disabling, offload/reload events, and memory adaptation. The trace shows that, under high request pressure, Nightjar disables speculative decoding and coordinates this transition with elastic memory management: it offloads the draft model when needed, expands the KV-cache budget, and admits more requests for concurrent processing. This elastic memory management helps Nightjar process the burst faster than static-memory baselines by allowing the system to serve more requests during the high-load phase. Detailed timelines are provided in Appendix~\ref{sec:appendix_stress}.


\subsubsection{Compare with different speculative length}
\begin{figure}[pos=h]
    \centering
    \includegraphics[width=0.9\linewidth]{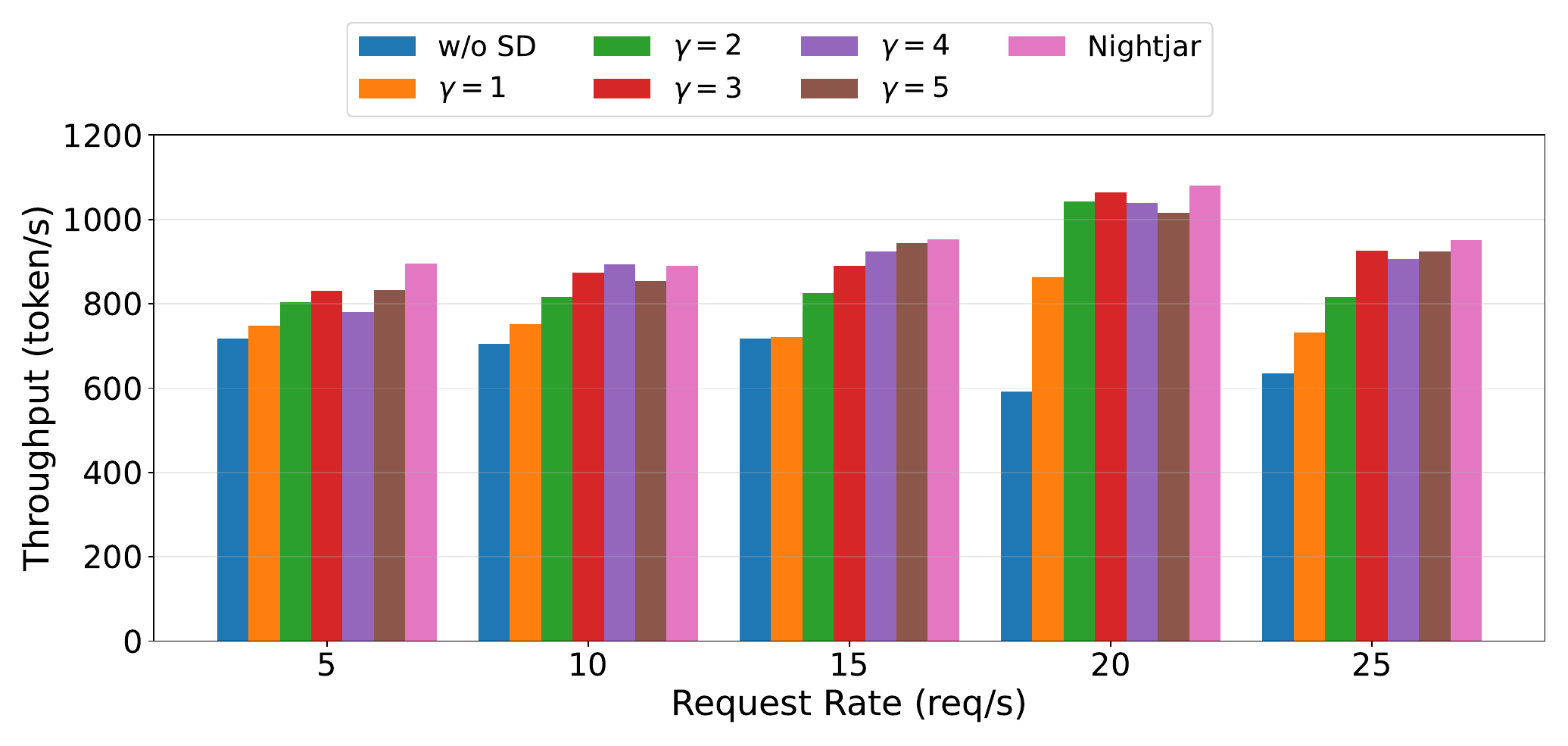}
    \caption{Throughput comparison for static request rate on SpecBench dataset. Comparison among different speculative length $\gamma$ for 13B model.}
    \label{fig:diff_length_comparison_multi_qps}
    \end{figure}
    As shown in Figure \ref{fig:diff_length_comparison_multi_qps}, when compared against the best-performing fixed speculative lengths (such as $\gamma=2$ or $\gamma=4$ at their peak efficiency) for 13B model,  Nightjar consistently maintains a 1\% to 16\% performance advantage. This demonstrates exceptional robustness, as Nightjar never falls behind the optimal static configuration.

\subsubsection{Scalability Analysis on Multi-GPU Settings}
\begin{figure}[pos=h]
    \centering
    \includegraphics[width=0.95\linewidth]{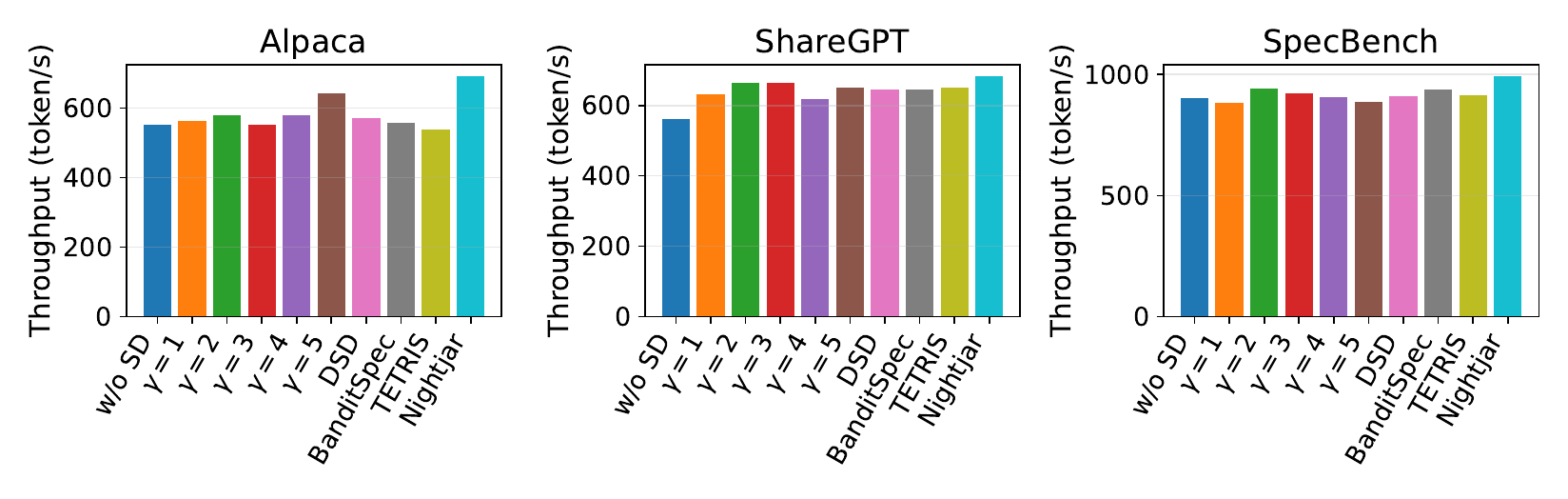}
    \caption{Performance comparison of 30B model.}
    \label{fig:dynamic_comparison_30B}
    \end{figure}

In this experiment, we deploy a 30B-scale model across two NVIDIA L20 (48GB) GPUs using tensor parallelism. Specifically, we use Qwen2.5-32B-Instruct as the target model and Qwen2.5-0.5B-Instruct as the draft model. As shown in Figure~\ref{fig:dynamic_comparison_30B}, by comparing Nightjar against other baselines, we observe that Nightjar maintains its performance advantage even when scaling to larger models distributed across multiple GPUs. Compared with strong adaptive baselines, Nightjar improves throughput by 20.7\% and 24.0\% over DSD and BanditSpec on Alpaca, 5.7\% and 5.6\% on ShareGPT, and 8.9\% and 5.8\% on SpecBench, respectively. It also consistently outperforms all fixed speculative lengths and Vanilla Decoding (w/o SD), showing that Nightjar can still effectively adapt speculative decoding decisions under the larger memory and multi-GPU execution setting. These results confirm that the proposed design generalizes well beyond single-node settings and remains effective for serving large models in distributed environments.

\subsubsection{Long-Context Evaluation}

\begin{table*}[pos=t]
\centering
\caption{Added 128K-context performance experiment on Llama-3.1-8B with draft model Llama-3.2-1B using two L20 48GB GPUs (TP=2), with 122880-token inputs and 16-token outputs. ``\#Prompts'' denotes the number of simultaneously injected long-context prompts.}
\label{tab:long_context_128k}
\small
\setlength{\tabcolsep}{5pt}
\begin{tabular}{c|c|c|c|c|c}
\hline
\textbf{\#Prompts (simultaneous)} & \textbf{Variant} & \textbf{Throughput} & \textbf{Mean TTFT} & \textbf{P99 E2E} & \textbf{Mean E2E} \\
\hline
\multirow{2}{*}{2} & Nightjar w/ offload & 2969.78 & 60219.14 & 82737.98 & 82666.96 \\
 & Nightjar (w/o offload) & 2788.81 & 60269.17 & 88106.86 & 88035.75 \\
\hline
\multirow{2}{*}{8} & Nightjar w/ offload & 2538.15 & 216403.01 & 387078.10 & 255369.77 \\
 & Nightjar (w/o offload) & 2327.35 & 217154.55 & 388760.71 & 259706.70 \\
\hline
\multirow{2}{*}{16} & Nightjar w/ offload & 2596.50 & 379297.19 & 723068.78 & 469455.77 \\
 & Nightjar (w/o offload) & 2516.59 & 414642.20 & 780591.13 & 462159.05 \\
\hline
\end{tabular}\par
\vspace{0.05cm}
{\small Throughput is reported in tokens/s; TTFT and E2E are reported in ms.}
\end{table*}

We further evaluate Nightjar in a KV-dominant long-context setting. As shown in Table~\ref{tab:long_context_128k}, the offload variant consistently improves throughput over the non-offload variant at 128K context. It also reduces mean E2E latency at 2 and 8 simultaneous prompts. At 16 prompts, throughput still improves, but mean E2E latency no longer decreases. These results indicate that, in the 128K setting, elastic offloading helps relieve KV-cache pressure, although the end-to-end latency benefit becomes limited at higher concurrency. Detailed memory-adaptation statistics are provided in Appendix~\ref{sec:appendix_long_context_memory}.

\subsubsection{Overhead of Elastic Memory Operations}

\begin{table}[h]
    \centering
    \caption{Overhead of elastic memory operations for 30B model on 2 L20 GPUs.}
    \label{tab:overhead}
    \begin{tabular}{lc}
    \toprule
    \textbf{Operation} & \textbf{Latency} \\
    \midrule
    KV cache block expansion & 143.9 ms \\
    KV cache block contraction (Triton-accelerated) & 11.9 ms \\
    Draft Model Reload Dispatch (CPU Overhead) & 21.9 $\mu$s \\
    \bottomrule
    \end{tabular}
    \end{table}
As shown in Table~\ref{tab:overhead}, we further quantified the overhead of these dynamic adjustments to ensure they do not disrupt real-time serving for 30B model on 2 L20 GPUs. The KV cache expansion process completes in 143.9 ms, which is acceptable because it is triggered only after sustained memory pressure and is used to reclaim draft-model memory for the KV cache during high-load phases. Conversely, KV cache contraction, accelerated by our custom Triton kernel, completes in just 11.9 ms, enabling efficient draft-model reload when the system transitions back to low-load states.
   The reported 21.9 $\mu$s cost is the CPU dispatch overhead for issuing an asynchronous reload. The PCIe transfer proceeds in the background, and the system continues serving requests with speculation disabled until the draft model is ready. We provide a more detailed discussion of transfer time under different bandwidth conditions in Appendix~\ref{sec:appendix_pcie}.

\begin{figure}[pos=h]
    \centering
    \includegraphics[width=0.7\linewidth]{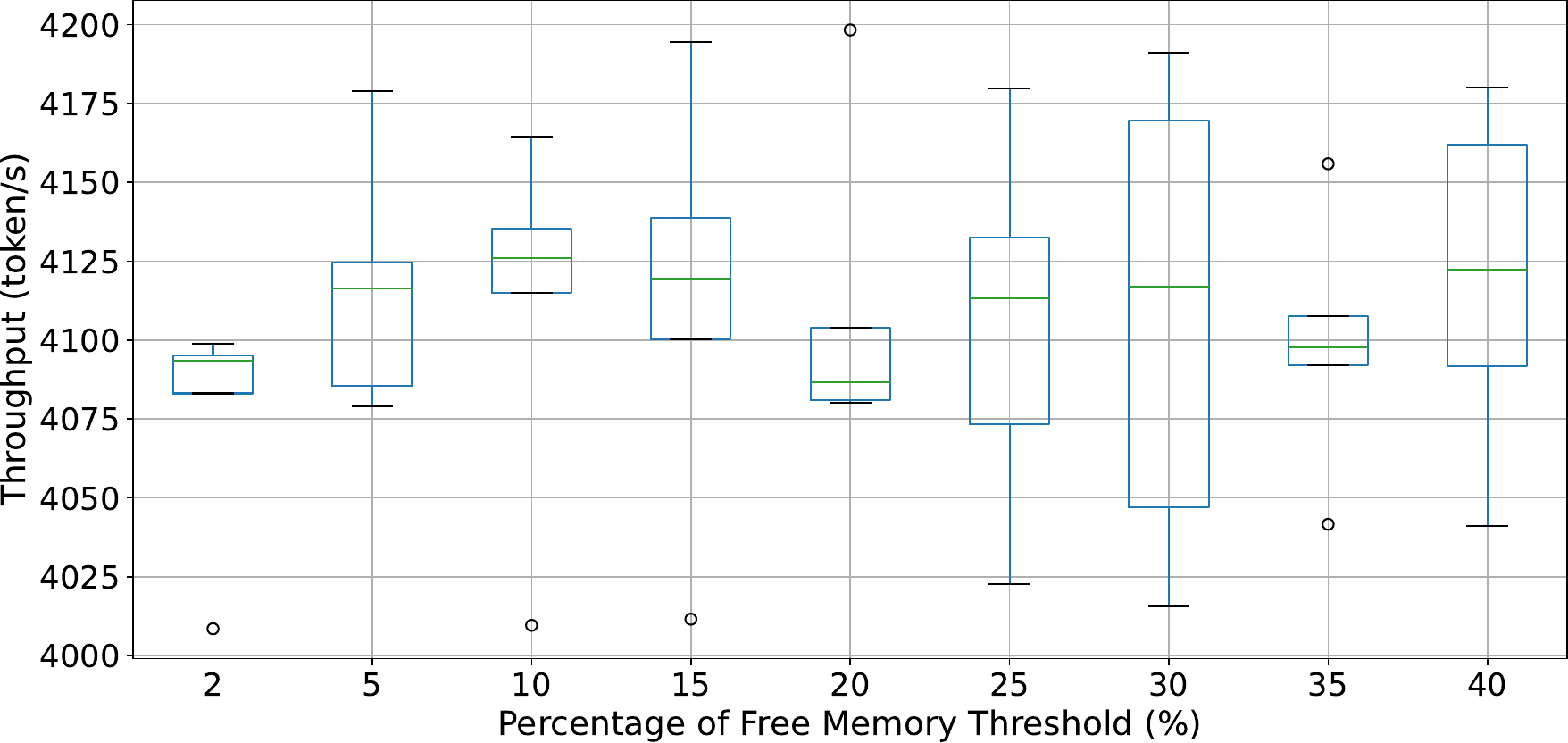}
    \caption{Performance comparison across different thresholds.}
    \label{fig:threshold_comparison}
    \end{figure}
\subsubsection{Sensitivity to Trigger Thresholds}
We further evaluate the sensitivity of system performance to the free KV cache threshold $\tau_{\text{low}}$, which serves as a key control parameter in the memory elasticity mechanism. Figure~\ref{fig:threshold_comparison} illustrates that the system reaches an optimal throughput plateau of approximately 6110 tok/s at a 10\% free cache threshold, which significantly exceeds the fixed baseline performance of 5768.12 tok/s. This result indicates that a 10\% buffer provides the ideal balance between sustaining speculative decoding and preventing memory exhaustion, ensuring robust QoS under high request load.

\subsubsection{Sensitivity to the Persistence Window}

\begin{table}[h]
    \centering
    \caption{Sensitivity of Nightjar to the persistence window $T_{persist}$.}
    \label{tab:tpersist_main}
    \small
    \setlength{\tabcolsep}{4pt}
    \begin{tabular}{c|ccc}
        \hline
        $T_{persist}$ & \shortstack{Throughput\\(tokens/s)} & \shortstack{Mean E2E\\Latency (ms)} & \shortstack{Offload\\Count} \\
        \hline
        1 & 1416.50 & 8419.28 & 5 \\
        2 & 1615.59 & 8119.68 & 2 \\
        3 & 1692.48 & 7620.77 & 1 \\
        5 & 1642.87 & 7958.10 & 1 \\
        8 & 1567.59 & 8729.64 & 1 \\
        \hline
    \end{tabular}
\end{table}

Table~\ref{tab:tpersist_main} shows how the persistence window affects switching behavior. When $T_{\text{persist}}$ is too small, the system reacts quickly but performs more offload/reload operations; when it is too large, the system responds more slowly to sustained memory pressure. $T_{\text{persist}}=3$ gives the best result in our setting, with the highest throughput (1692.48 tokens/s), the lowest mean E2E latency (7620.77 ms), and only one offload. We therefore use $T_{\text{persist}}=3$ as the default setting.

\section{Related Work}
\subsection{Speculative Decoding in LLM Serving}

While various approaches have sought to optimize speculative decoding, they often fail to address the dynamic and unpredictable nature of real-world LLM serving loads. Initial efforts focused on single-request optimization \cite{wang_opt-tree_2025,zhang_adaeagle_2024,zhang_draft_2024,brown_dynamic_2024,huang_specdec_2024} or static offline batches \cite{su2023synergy, houbanditspec}, employing predictive algorithms \cite{zhang_adaeagle_2024} or confidence-based methods \cite{huang_specserve_2025, wang_opt-tree_2025,wu2025tetris}, but their static policies cannot handle the changing request rates in real-world environments. More recent works have explored dynamic speculation length adjustment in serving systems \cite{su2023synergy,liu2024optimizing,huang_specserve_2025,li_adaserve_2025}, yet they exhibit significant limitations. For instance, SpecServe \cite{huang_specserve_2025} and TETRIS \cite{wu2025tetris} rely on draft confidence scores, but this can lead to poor performance with large batch sizes and still requires generating at least one token. DSD \cite{liu2024optimizing} optimizes length via linear goodput modeling, but its reliance on past average acceptance rates introduces a critical vulnerability. Because these methods depend on historical speculative observations, disabling speculation also stops the collection of fresh acceptance statistics, which can make later reactivation less reliable. Our exploration-exploitation approach is designed to mitigate this risk. Although BanditSpec \cite{houbanditspec} introduces a Multi-Armed Bandit paradigm, it is constrained by its static design, which is incompatible with the continuous batching \cite{kwon2023efficient} of modern serving systems.
In addition, current methods usually ignore the “switching cost” incurred when re-enabling speculative decoding from a length of 0, which is the overhead of KV cache reconstruction for the draft model.
\subsection{Unified Memory Management for Weights and KV cache in LLM Serving}
Oneiros \cite{li2025oneiros} targets multi-model deployment by introducing a dynamic memory remapping engine that redirects GPU memory reserved for model parameters to the KV cache. Unlike Nightjar, which completely offloads draft model weights when they are not in use, Oneiros operates on active models. By performing remapping at layer granularity, it enables models to remain in a serving state while utilizing asynchronous transfers to fetch required parameters in real-time before execution. 
eLLM \cite{xu2025ellm} focuses on intra-model efficiency, enabling the KV cache and activation tensors to borrow space from a shared physical memory pool, effectively breaking the traditional isolation between these resources.
FineServe \cite{bin2025fineserve} provides a unified management framework for both weights and KV cache across multiple models, specifically addressing the challenges of serving models with heterogeneous quantization precisions. These works primarily optimize memory sharing among continuously active models or tensors. In contrast, Nightjar is designed specifically for speculative decoding, where draft model residency and KV cache capacity compete for GPU memory across decoding modes. By offloading the draft model during non-speculative phases, Nightjar enlarges the KV cache block pool under high load, enabling higher achievable concurrency rather than merely redistributing memory among active components.
Nightjar is complementary to recent hardware-software co-design efforts for efficient LLM inference \cite{liu2026isscc_llm_accel}, while our contribution focuses on serving-time speculative control and memory adaptation under dynamic workloads. Related system-level work also includes Teegraph \cite{fu2022teegraph} and Subtraction of Hyperledger Fabric \cite{fu2024subtraction}, which likewise emphasize practical system design under resource constraints.

\section{Conclusion and Future Work}

In this paper, we presented \textit{Nightjar}, a dynamic adaptive framework for speculative decoding in real-world LLM serving. We show that speculative decoding is not universally beneficial: while it improves performance under low load, its verification overhead and draft model memory footprint can degrade efficiency under high load. Nightjar addresses this by jointly optimizing speculative length selection and GPU memory allocation, dynamically disabling speculation and offloading the draft model to expand KV cache capacity when necessary. A current limitation is that the switching-cost estimates are obtained from offline profiling and queried through a deployment-specific lookup table. Future work will focus on online calibration for new hardware environments and on extending Nightjar to multi-GPU and distributed serving. Overall, our results highlight the importance of coordinating speculative policies with memory resource management under dynamic workloads.

\appendix

\section{Appendix A: Detailed Theoretical Analysis of Regret Bound}
\label{sec:appendix_proof}

In this appendix, we provide the complete derivation of the regret bound for the Nightjar algorithm. We demonstrate that the hierarchical bin-based structure effectively controls the switching cost, ensuring sublinear regret. Our objective is to minimize the cumulative latency per token. A key aspect of our formulation is the specific structure of the loss function, which penalizes strategy switches inversely proportional to the speculative length.

\subsection{Assumptions}

We analyze the regret under the standard stochastic setting.
For each batch size $B$, and each arm $\gamma \in \{0,\dots,\Gamma_{\max}\}$, the per-step latency $\ell_t(\gamma)$ is drawn i.i.d. from a stationary distribution with mean $\mu_{B,\gamma}$ whenever $B_t=B$.
We assume bounded losses: $\ell_t(\gamma)\in[0,\ell_{\max}]$.
The switching-cost lookup table provides a uniform upper bound $C_{\text{switch}}(\delta_{\max,t},B_t)\le C_{\max}$ for all feasible $(\delta_{\max,t},B_t)$.

\subsection{Regret Decomposition}

Recall the loss
\begin{equation}
    L_t(\gamma_t)=\ell_t(\gamma_t)+\mathbb{I}(\gamma_{t-1}=0\land \gamma_t>0)\cdot \frac{C_{\text{switch}}(\delta_{\max,t},B_t)}{\gamma_t}.
\end{equation}
The benchmark is the best fixed arm per batch size:
$\gamma^*(B)=\arg\min_{\gamma}\mu_{B,\gamma}$.
Thus the cumulative regret is
\begin{equation}
    \begin{aligned}
    R(T)
    &= \sum_{t=1}^T 
       \big(\ell_t(\gamma_t)
       - \ell_t(\gamma^*(B_t))\big) \\
    &\quad
       + \sum_{t=1}^T 
       \mathbb{I}(\text{switch}_t)
       \cdot 
       \frac{C_{\text{switch}}(\delta_{\max,t},B_t)}{\gamma_t} \\
    &\triangleq 
       R_{\text{perf}}(T)
       + R_{\text{switch}}(T).
    \end{aligned}
    \end{equation}

\subsection{Bounding the Switching Regret}

Since any valid switch implies $\gamma_t \ge 1$ and the switching cost is bounded by $C_{\text{switch}}(\delta_{\max,t},B_t) \le C_{\max}$, it follows that:
\begin{equation}
\frac{C_{\text{switch}}(\delta_{\max,t},B_t)}{\gamma_t} \le C_{\max}.
\end{equation}
By bin locking, a switch can only occur at bin boundaries.
Let $T_B=\sum_{t=1}^T\mathbb{I}(B_t=B)$ be the number of steps with batch size $B$.
The Nightjar schedule for each $B$ partitions its own timeline into blocks of length $H_B=2^{j_B-1}$, each containing $\Theta(\sqrt{H_B})$ bins.
Hence the number of bins (and thus switches) on the $B$-timeline is $O(\sqrt{T_B})$.
Summing over all batch sizes yields
\begin{equation}
    \begin{aligned}
    S_T 
    &\le \sum_{B=1}^{B_{\max}} O(\sqrt{T_B}) \\
    &\le O(\sqrt{B_{\max}T}),
    \end{aligned}
    \end{equation}
    
    \begin{equation}
    R_{\text{switch}}(T)
    \le C_{\max} S_T
    = \tilde{O}(\sqrt{T}).
    \end{equation}
treating $B_{\max}$ as a constant.

\subsection{Bounding the Performance Regret}

Within each $B$-timeline, Nightjar follows the ADA-BINGREEDY schedule: in bin $b_B$ it explores with probability $p_{b_B}=1/b_B$, otherwise exploits using the empirically best arm.
In a block with $K_B=\Theta(\sqrt{H_B})$ bins, the expected number of exploration bins is
$\sum_{b_B=1}^{K_B}\frac{1}{b_B}=O(\log K_B)$.
Each bin lasts $\Theta(\sqrt{H_B})$ steps and incurs per-step regret at most $\ell_{\max}$, so the exploration regret per block is
$O(\ell_{\max}\sqrt{H_B}\log H_B)$.
Summing over geometrically increasing blocks up to horizon $T_B$ gives
$R_{\text{explore}}^{(B)}(T_B)=\tilde{O}(\sqrt{T_B})$.

For exploitation bins, since $\ell_t(\gamma)\in[0,\ell_{\max}]$ and samples are i.i.d. with mean $\mu_{B,\gamma}$, Hoeffding's inequality implies that the probability of choosing a suboptimal arm decays exponentially with the number of samples.
Following the analysis of ADA-BINGREEDY~\cite{luo2018efficient}, the exploitation regret is dominated by the exploration term, yielding
$R_{\text{perf}}^{(B)}(T_B)=\tilde{O}(\sqrt{T_B})$.
Summing over $B$ and using $\sum_B \sqrt{T_B}\le \sqrt{B_{\max}T}$, we obtain
$R_{\text{perf}}(T)=\tilde{O}(\sqrt{T})$.

\subsection{Conclusion}

Combining $R_{\text{switch}}(T)=\tilde{O}(\sqrt{T})$ and $R_{\text{perf}}(T)=\tilde{O}(\sqrt{T})$ completes the proof:
\begin{equation}
R(T)=\tilde{O}(\sqrt{T}).
\end{equation}

\section{Appendix B: Model-Free Drafting Control}
\label{sec:appendix_ngram}

\begin{table*}[htbp]
\centering
\caption{Lightweight control experiment for model-free N-gram drafting on ShareGPT using DeepSeek-R1-Distill-Qwen-7B.}
\label{tab:appendix_ngram}
\scriptsize
\setlength{\tabcolsep}{4pt}
\begin{tabular}{c|cc|cc|cc|cc|cc}
\hline
\textbf{Req/s} & \multicolumn{2}{c|}{\textbf{Nightjar with N-gram ($\gamma_{\max}=5$)}} & \multicolumn{2}{c|}{\textbf{Fixed N-gram ($\gamma=1$)}} & \multicolumn{2}{c|}{\textbf{Fixed N-gram ($\gamma=3$)}} & \multicolumn{2}{c|}{\textbf{Fixed N-gram ($\gamma=5$)}} & \multicolumn{2}{c}{\textbf{Vanilla Decoding (w/o SD)}} \\
 & Throughput & Latency & Throughput & Latency & Throughput & Latency & Throughput & Latency & Throughput & Latency \\
\hline
1 & 359.26 & 3352.75 & 347.90 & 3482.04 & 194.19 & 3386.52 & 194.29 & 3438.75 & 194.69 & 3422.72 \\
3 & 380.46 & 3770.04 & 374.79 & 3804.87 & 375.46 & 3805.02 & 374.77 & 3825.88 & 366.29 & 3751.15 \\
5 & 461.92 & 3984.00 & 449.91 & 4081.49 & 451.55 & 4078.01 & 454.25 & 4006.54 & 446.07 & 4045.15 \\
\hline
\end{tabular}
\end{table*}

Table~\ref{tab:appendix_ngram} shows that Nightjar's adaptive speculative-length selection also generalizes to a model-free draft setting. In this control experiment there is no neural draft model and thus no draft-model offload/reload mechanism, so the gain comes only from adaptive speculative-length selection.

\section{Appendix C: Bursty Stress-Test Details}
\label{sec:appendix_stress}

\begin{figure*}[htbp]
\centering
\includegraphics[width=\linewidth]{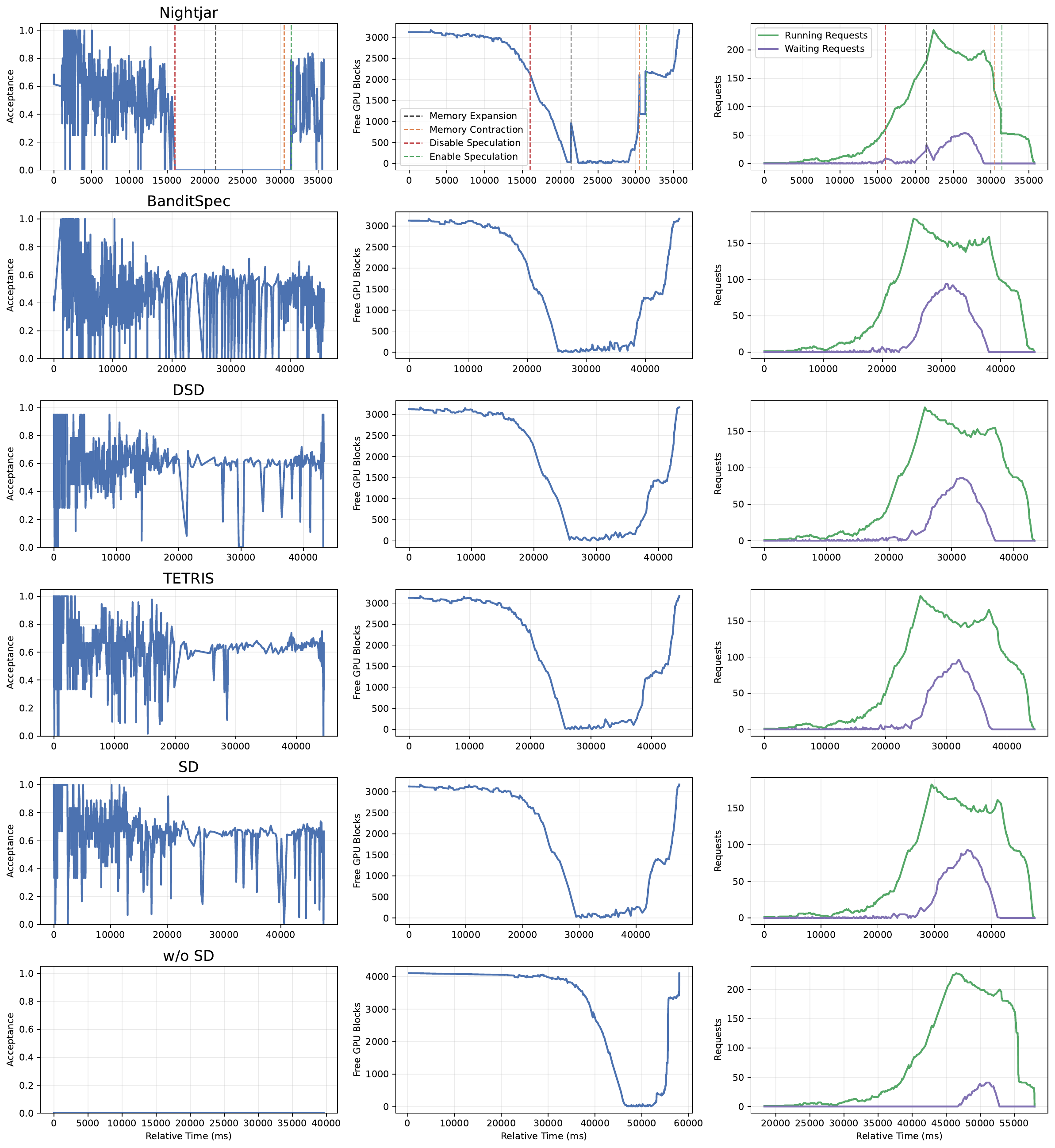}
\caption{Burst-spike stress-test timeline showing when speculation is disabled. Nightjar delays entry into the disabled state compared with the static-memory baseline.}
\label{fig:appendix_stress_disable_timeline}
\end{figure*}

\begin{table*}[htbp]
\centering
\caption{Stress-test comparison under three workload traces: Burst spike (mostly moderate load with short sharp bursts), High/low oscillation (alternating sustained high-load and low-load phases), and Synchronized multi-peak (multiple request streams peaking at similar times).}
\label{tab:appendix_stress}
\small
\begin{tabular}{c|c|c|c|c}
\hline
Trace & Variant & Total throughput (tokens/s) & Mean E2E (ms) & P99 E2E (ms) \\
\hline
\multirow{2}{*}{Burst spike} & Nightjar & 1934.54 & 12287.56 & 16146.50 \\
 & Nightjar (w/o offload) & 1793.88 & 12294.09 & 17238.36 \\
\hline
\multirow{2}{*}{High/low oscillation} & Nightjar & 1394.27 & 6479.80 & 10147.85 \\
 & Nightjar (w/o offload) & 1259.86 & 7567.89 & 11007.64 \\
\hline
\multirow{2}{*}{Synchronized multi-peak} & Nightjar & 724.47 & 6537.22 & 9690.76 \\
 & Nightjar (w/o offload) & 707.28 & 8537.36 & 12829.96 \\
\hline
\end{tabular}
\end{table*}

\begin{table*}[htbp]
\centering
\caption{Memory adaptation and KV-migration timing in the stress traces.}
\label{tab:appendix_stress_timing}
\small
\resizebox{\textwidth}{!}{
\begin{tabular}{l|c|c|c|c|c|c}
\hline
Trace & Adapt. cycles & KV mig. events & Moved blocks & Affected seqs & KV mig. ms & Expand / contract ms \\
\hline
Burst spike & 1 & 1 & 292 & 49 & 12.15 & 88.95 / 306.00 \\
High/low oscillation & 1 & 1 & 57 & 42 & 2.48 & 10.01 / 32.70 \\
Synchronized multi-peak & 3 & 3 & 0 & 0 & 3.02 / 2.84 / 0.90 & 13.70 / 9.86; 9.88 / 9.98; 5.08 / 3.26 \\
\hline
\end{tabular}}
\end{table*}

Figure~\ref{fig:appendix_stress_disable_timeline} further shows when speculation is disabled during the burst-spike trace. With elastic memory management, speculation is disabled only after the Memory Manager has already attempted adaptation; in the corresponding static-memory baseline, disabled intervals appear earlier and persist for more steps. This confirms that Nightjar does not keep speculation enabled unconditionally, but delays and shortens disabled intervals through elastic memory management. Tables~\ref{tab:appendix_stress} and~\ref{tab:appendix_stress_timing} further confirm that bursty arrivals can trigger real memory adaptation and nonzero KV migration, but the migration stage remains bounded relative to end-to-end request latency. Even in the burst-spike trace, where 292 KV blocks are moved and 49 sequences are affected, the recorded KV-migration stage is 12.15 ms.

\section{Appendix D: Long-Context Memory Adaptation Details}
\label{sec:appendix_long_context_memory}

\begin{table}[H]
\centering
\caption{Memory-adaptation evidence from the 128K-context traces, with measured KV-cache expansion/contraction counts and elapsed times.}
\label{tab:long_context_128k_memory}
\small
\setlength{\tabcolsep}{3pt}
\resizebox{0.92\columnwidth}{!}{
\begin{tabular}{c|c|c|c}
\hline
\textbf{\#Prompts} & \textbf{Variant} & \makecell{\textbf{KV expand} \\ \textbf{count / time}} & \makecell{\textbf{KV contract} \\ \textbf{count / time}} \\
\hline
2 & Nightjar w/ offload  & 1 / 553.51 ms & 1 / 659.87 ms \\
\hline
8 & Nightjar w/ offload & 1 / 571.87 ms & 1 / 844.28 ms \\
\hline
16 & Nightjar w/ offload  & 1 / 574.95 ms & 1 / 890.05 ms \\
\hline
\end{tabular}
}
\end{table}

The results in Table~\ref{tab:long_context_128k_memory} show that Nightjar's elastic memory management remains effective in a KV-dominant long-context setting. Across all three runs, the system completes one expansion and one contraction cycle, and the contraction stage remains bounded. This indicates that longer contexts increase KV pressure and can delay draft-model reload, but in the evaluated traces they do not necessarily force large KV migration during contraction.

\section{Appendix E: PCIe Transfer Sensitivity}
\label{sec:appendix_pcie}

\begin{table*}[!b]
\centering
\caption{Background completion time for draft-model offload/reload under different effective PCIe bandwidths. We use a 942.511 MiB BF16 0.5B draft model; the reported value is the wall-clock time until all concurrent transfers finish, not the CPU-side dispatch overhead.}
\label{tab:appendix_pcie}
\small
\begin{tabular}{c|c|c|c|c|c|c}
\hline
Transfer concurrency & Total transferred data & 12 GiB/s & 16 GiB/s & 23 GiB/s & 32 GiB/s & 64 GiB/s \\
\hline
1 & 0.92 GiB & 76.70 ms & 57.53 ms & 40.02 ms & 28.76 ms & 14.38 ms \\
2 & 1.84 GiB & 153.40 ms & 115.05 ms & 80.04 ms & 57.53 ms & 28.76 ms \\
4 & 3.68 GiB & 306.80 ms & 230.10 ms & 160.07 ms & 115.05 ms & 57.53 ms \\
8 & 7.36 GiB & 613.61 ms & 460.21 ms & 320.15 ms & 230.10 ms & 115.05 ms \\
\hline
\end{tabular}
\end{table*}

Table~\ref{tab:appendix_pcie} makes the transfer-level assumption explicit using the measured BF16 weight payload of the 0.5B draft model, 942.511 MiB (approximately 0.92 GiB). The 21.9 $\mu$s number in the main text is only the CPU-side dispatch overhead; the transfers themselves can take tens to hundreds of milliseconds depending on PCIe bandwidth and concurrency. For example, the single-transfer completion time at 23 GiB/s is 40.02 ms, while 8 concurrent transfers increase it to 320.15 ms, and the same 8-transfer case reaches 613.61 ms at 12 GiB/s. This is why Nightjar treats reload as a background transfer rather than something that blocks request execution.

\section{Appendix F: 13B Latency-Breakdown Case Study}
\label{sec:appendix_13b_case}

\begin{table*}[!t]
\centering
\caption{Latency breakdown on Vicuna-13B for the representative cases where Nightjar trades a small latency increase for higher throughput.}
\label{tab:appendix_13b_case}
\small
\begin{tabular}{c|c|c|c|c|c}
\hline
Dataset & Method & Throughput (tokens/s) & Mean TTFT (ms) & Mean TPOT (ms) & Mean E2E latency (ms) \\
\hline
\multirow{3}{*}{Alpaca} & DSD & 603.32 & 248.59 & 49.23 & \textbf{21212.44} \\
 & TETRIS & 702.65 & 538.02 & 61.18 & 22133.43 \\
 & Nightjar & \textbf{721.23} & 371.29 & 50.04 & 21525.49 \\
\hline
\multirow{3}{*}{ShareGPT} & DSD & 1244.96 & 2913.21 & 52.05 & 10057.72 \\
 & TETRIS & 1628.86 & 1658.82 & 55.67 & \textbf{8046.25} \\
 & Nightjar & \textbf{1729.68} & 2113.21 & 55.65 & 8133.92 \\
\hline
\end{tabular}
\end{table*}

Table~\ref{tab:appendix_13b_case} shows that the small end-to-end latency penalty mainly comes from front-end waiting time rather than slower per-token decoding. The TPOT values are nearly identical across the compared methods. In other words, on Vicuna-13B the throughput gain mainly comes from serving more requests with larger or more efficient decoding batches, which can leave some requests waiting longer in the queue and therefore increases TTFT slightly.

\section{Declaration of generative AI and AI-assisted technologies in the manuscript preparation process}
During the preparation of this work the authors used Gemini in order to polish the manuscript. After using this tool/service, the authors reviewed and edited the content as needed and take full responsibility for the content of the published article.
\printcredits



\bibliographystyle{cas-model2-names}

\bibliography{custom}

\end{document}